\documentclass[aps,prb,nofootinbib,reprint,twocolumn,longbibliography,superscriptaddress,showpacs]{revtex4-1}
\usepackage{amsmath}
\usepackage{amsthm}
\usepackage{amsfonts}
\usepackage{comment}
\usepackage{hyperref}
\usepackage{upgreek}
\usepackage{graphicx}
\usepackage{xcolor}

\newcommand{\bchi}{\boldsymbol{\upchi}}
\newcommand{\bxi}{\boldsymbol{\upxi}}
\newtheorem{lemma}{Lemma}

\newcommand{\eqnref}[1]{Eq.\ (\ref{#1})}
\newcommand{\figref}[1]{Fig.\ \ref{#1}}
\newcommand{\secref}[1]{Sec.\ \ref{#1}}

\newcommand{\bra}[1]{\langle #1|}
\newcommand{\ket}[1]{|#1\rangle}

\newcommand{\Tr}{\mathrm{Tr}}
\newcommand{\SO}{\mathrm{SO}}
\newcommand{\SU}{\mathrm{SU}}
\newcommand{\U}{\mathrm{U}}
\newcommand{\PSU}{\mathrm{PSU}}

\begin{document}

\author{Dominic V. Else}
\affiliation{Centre for Engineered Quantum Systems, School of Physics,
The University of Sydney, Sydney, NSW 2006, Australia}
\affiliation{Department of Physics, University of California, Santa Barbara, CA
93106, USA}
\author{Stephen D. Bartlett}
\affiliation{Centre for Engineered Quantum Systems, School of Physics,
The University of Sydney, Sydney, NSW 2006, Australia}
\author{Andrew C. Doherty}
\affiliation{Centre for Engineered Quantum Systems, School of Physics,
The University of Sydney, Sydney, NSW 2006, Australia}

\title{The hidden symmetry-breaking picture of symmetry-protected topological
order}

\begin{abstract}
We generalize the hidden symmetry-breaking picture of
symmetry-protected topological (SPT) order developed by Kennedy and Tasaki in
the context of the Haldane phase.  Our generalization applies to a wide class of
SPT phases in one-dimensional spin chains, protected by an on-site
representation of a finite abelian group.  This generalization takes the form of
a non-local unitary map that relates local symmetry-respecting Hamiltonians in
an SPT phase to local Hamiltonians in a symmetry-broken phase.  Using this
unitary, we establish a relation between the two-point correlation functions
that characterize fully symmetry-broken phases with the string-order correlation
functions that characterise the SPT phases, therefore establishing the
perspective in these systems that SPT phases are characterised by \emph{hidden}
symmetry-breaking.  Our generalization is also applied to systems with
continuous symmetries, including $\SO(2k+1)$ and $\SU(k)$.
\end{abstract}


\maketitle

\section{Introduction}
In the traditional Landau paradigm, order in condensed matter systems is viewed
as synonymous with the spontaneous breaking of a symmetry. However, it is now
well-established that at zero temperature there exist topologically ordered phases, such as those of
the fractional quantum Hall effect, which cannot be understood
through the Landau paradigm. Additionally, one can also consider
\emph{symmetry-protected topological (SPT)} phases~\cite{wen_lu}, which are not topologically ordered
in the conventional sense, yet remain distinct from the trivial phase in
the presence of an appropriate symmetry. A well-known example of an SPT phase is
the Haldane phase of antiferromagnetic spin-1 chains,  which is protected by
the $D_2 \cong Z_2 \times Z_2$ symmetry comprising $\pi$ rotations about a
set of orthogonal axes.
The ``topological'' nature
of the Haldane phase is manifested in a number of ways, such as the long-range string
order~\cite{aklt_string_order}, fractionalized
edge modes~\cite{haldane_emergent_edge1}, degenerate entanglement spectrum~\cite{pollmann-prb-2010}, and long-range localizable
entanglement~\cite{diverging_prl,camposvenuti}.  It is now known that many of
the interesting properties of the Haldane phase extend in general to
SPT phases of one-dimensional spin chains protected by a unitary ``on-site'' representation of an arbitrary
symmetry group $G$ (for which the Haldane phase, with $G = Z_2 \times Z_2$, is the simplest
non-trivial example), which have been fully classified~\cite{chen_gu_wen,schuch}.

An early and influential characterization of the Haldane phase was provided by
Kennedy and Tasaki~\cite{kt,*kt2} (see also Ref.~\onlinecite{pollmann-arxiv-2009}). They constructed a non-local unitary [which we refer to as
the \emph{Kennedy-Tasaki (KT) transformation}] to transform the Haldane phase to a
conventional symmetry-breaking phase. Although the transformation is non-local,
for any local
Hamiltonian $H$ that respects the $D_2$ rotation symmetry, the KT
transformation yields another $D_2$-symmetric local Hamiltonian
$\widetilde{H}$. It turns out that if $H$ is in the SPT Haldane phase, then
$\widetilde{H}$ will have a space of four degenerate locally-distinguishable
ground states corresponding to the spontaneous breaking of the $D_2$ rotation
symmetry, i.e.\ $\widetilde{H}$ is in the maximal symmetry-breaking phase for the $D_2$
symmetry.
Thus, the ordering in the SPT phase is interpreted as ``hidden symmetry-breaking''.
Furthermore, the long-ranged string order in the Haldane phase is related by the KT transformation to
conventional long-ranged order in the symmetry breaking phase.

The Haldane phase is also a special case of the SPT phases that were classified
in Refs.~\onlinecite{chen_gu_wen,schuch} through the symmetry properties of a matrix-product state ansatz for
the ground state. The general result is that the distinct SPT phases with
respect to an on-site representation of a symmetry group $G$ are classified by
the second cohomology classes of the projective representations of the symmetry
group. This classification appears to be the most fundamental and general approach to SPT
phases for one-dimensional chains with an on-site symmetry. 
However, the connection with the
original
hidden symmetry-breaking picture
of Kennedy and Tasaki has not been explored. 
As a result, it remains
unclear to what extent other SPT phases can also be understood to arise from a similar
hidden symmetry-breaking mechanism (although see Ref.~\onlinecite{SO_string_order} for one example).

In this paper, we extend the hidden-symmetry breaking picture to any SPT phase
protected in a one-dimensional spin chain by an on-site representation of a
finite abelian group,
provided that
the cohomology class describing the phase satisfies a condition called
\emph{maximal non-commutativity}. (We say that a cohomology class is maximally
non-commutative if, in the corresponding projective representations, for the matrix
representation of any non-trivial group element there exists at least one other
matrix in the representation with which it does not commute~\cite{else_schwarz_bartlett_doherty_symmetry}.)  We achieve this by constructing a
suitable generalization of the KT transformation (presented in Sec.~\ref{sec_generalized_kt}), expressed explicitly
in terms of the appropriate cohomology class of the symmetry group, to
transform the SPT phase into a conventional symmetry-breaking phase.
The generalized KT transformation is essentially equivalent to the duality
transformation introduced by us in the context of quantum
computation \cite{big_paper}, and some of its properties were already discussed in the appendices
of that paper; however, our treatment here will be self-contained.  

Where it can be applied, our generalized KT transformation affords a
different perspective on properties of the SPT phase. For example, 
for abelian symmetry groups, SPT phases (and indeed, all
symmetric phases) can be identified from a pattern of string order (as we will
show, based in part on the results of Ref.~\onlinecite{detection_string_order}). For an SPT phase corresponding
to a maximally non-commutative cohomology class, this pattern of string order can be understood
in a natural way
through the generalized KT transformation, which relates it to the long-range
order characterizing the symmetry-breaking phase (just as the original KT
transformation does in the case of the Haldane phase).  We explore this perspective in Sec.~\ref{sec_string_order}. 
We remark that, although we are only able to consider finite abelian symmetry groups, these
groups can arise as subgroups for systems with a larger symmetry.
For SPT phases in systems that have a $\SO(2k+1)$ or $\SU(k)$ symmetry, 
in Sec.~\ref{sec:examples} we will exhibit an appropriate finite abelian subgroup that allows
the hidden symmetry breaking to be identified.  

We note that a closely related investigation, Ref.~\onlinecite{duivenvoorden_quella}, appeared shortly after our work and contains similar results to ours.  Where relevant, we will remark on some of the similarities and differences between the two works.  In particular, our generalized KT transformation
coincides with that of Ref. \onlinecite{duivenvoorden_quella} for the specific case of a $Z_N \times Z_N$ symmetry group.  

\section{The Kennedy-Tasaki transformation}
\label{sec:KTtrans}
Let us recall the definition of the unitary $\mathcal{D}_{KT}$ that effects the Kennedy-Tasaki transformation
for a chain of
$N$ spin-1's with open boundary conditions. It can be written as
\cite{kt_arbitrary_spin}
\begin{equation}
\label{DKT}
\mathcal{D}_{KT} = \prod_{j < k} \exp(i \pi S^z_j S^x_k),
\end{equation}
where $S^a_j$ ($a = x,y,z$) denotes the appropriate spin component operator for
the $j$-th spin.
This unitary is non-local, but, for any local observable $A$ that respects the
$D_2$ symmetry operations $\prod_j \exp(i\pi S^a_j)$ ($a = x,y,z$),
the transformed observable $\mathcal{D}_{KT} A \mathcal{D}_{KT}^{\dagger}$ remains local
and symmetry-respecting. Therefore, for any Hamiltonian $H$ that is the sum of
local symmetry-respecting interactions, one can generate the dual Hamiltonian
$\widetilde{H} = \mathcal{D}_{KT} H \mathcal{D}_{KT}^{\dagger}$. If $H$ is in the SPT
phase with respect to the $D_2$ symmetry, then $\widetilde{H}$ is expected to be in a conventional symmetry-breaking phase with respect to
the symmetry, with the four-fold degenerate edge states mapping under $\mathcal{D}_{KT}$ to the four locally distinguishable symmetry-breaking ground states.

In order to see that $\mathcal{D}_{KT}$ is a special case of the generalized KT
transformation to be defined later, we will want to express
$\mathcal{D}_{KT}$ in terms of the single-site basis $\{ \ket{x}, \ket{y}, \ket{z} \}$
(where $\ket{a}$ is the zero eigenstate of $S^a$ for $a = x,y,z$), that is the
simultaneous eigenbasis of the on-site representation of the symmetry.
Observe that
\begin{align}
\exp(i\pi S_z \otimes S_x) (\ket{a_1} \otimes \ket{a_2}) = (-1)^{\mu(a_1)
\nu(a_2)}
\ket{a_1} \otimes \ket{a_2},
\end{align}
where $\mu(a) = 1-\delta_{a,z}$ and $\nu(a) = 1-\delta_{a,x}$.
Hence we can write
\begin{equation}
\label{DKT_basis}
\mathcal{D}_{KT} = \sum_{a_1, \ldots, a_N} (-1)^{\sum_{j < k} \mu(a_j) \nu(a_k)}
\ket{a_1, \ldots, a_N} \bra{a_1, \ldots, a_N}.
\end{equation}

\section{Classification of SPT phases by cohomology classes}
Throughout this paper, we will assume a chain of $N$ spins, such that the
Hamiltonian commutes with the on-site representation $[u(g)]^{\otimes N}$
of a symmetry group $G$.  We will assume open boundary
conditions unless otherwise stated.  According to the general classification of SPT phases for on-site
symmetries in one-dimensional systems \cite{chen_gu_wen,schuch},
the different SPT phases for this symmetry can be classified by the second
cohomology group $H^2(G, \U(1))$, which is related to the projective
representations $V(g)$ for the group $G$, as we now describe. (One interpretation
of these projective representations is that they describe the action of the
symmetry on the fractionalized edge mode associated with each edge for open
boundary conditions; we note that there are several subtleties with the formal treatment of such edge modes, and we refer the reader to 
Sec.~3.1 of Ref.~\onlinecite{big_paper} for a careful discussion.\color{black})  By definition, a
projective representation $V(g)$ must satisfy
\begin{equation}
V(g_1) V(g_2) = \omega(g_1, g_2) V(g_1 g_2), \quad \forall g_1, g_2 \in G,
\end{equation}
where $\omega$ is a function mapping pairs of group elements to complex phase
factors, known as the \emph{factor system} of the projective representation. The associativity of matrix
multiplication implies that the factor system must satisfy the 2-cocycle condition
\begin{multline}
\label{cocycle_condn}
\omega(g_1, g_2) \omega(g_1 g_2, g_3) = \omega(g_2, g_3) \omega(g_1, g_2 g_3), \\ \forall g_1, g_2, g_3 \in G.
\end{multline}
Conversely, any $\omega$ satisfying \eqnref{cocycle_condn} is the factor system
for some projective representation \cite{berkovich_characters}. Furthermore,
given any projective representation, it is trivial to generate another one by
rephasing of the operators $V(g)$, i.e. $V(g) \to \beta(g) V(g)$, where $\beta$
is a function that sends group elements to phase factors. The effect on the
factor system is
\begin{equation}
\label{rephasing}
\omega(g_1, g_2) \to \beta(g_1 g_2)^{-1} \beta(g_1) \beta(g_2) \omega(g_1, g_2).
\end{equation}
Two factor systems related by a transformation of the form \eqnref{rephasing} are said to
be in the same \emph{cohomology class}, and the second cohomology group
$H^2(G,\U(1))$ comprises all the distinct cohomology classes for the group $G$.
We will denote by $[\omega]$ the cohomology class containing a given factor
system $\omega$.

In the case of the Haldane phase for spin-1 chains, the relevant symmetry group
is $D_2 = \{ 1, x, y, z \}$ (where $y = xz$), with
the on-site representation $u(a) = \exp(i \pi S_a)$ (for $a
= x,y,z$).
The Haldane phase
corresponds to the unique non-trivial cohomology class for the symmetry group $D_2$. We
can specify a representative factor system $\omega$ for this cohomology class by
giving an example of a projective representation for which $\omega$ is the
factor system, namely
\begin{equation}
\label{D2proj}
V(1) = \mathbb{I}, \quad V(x) = \sigma_x, \quad V(z) = \sigma_z, \quad V(y) = \sigma_x \sigma_z,
\end{equation}
where $\sigma_x$ and $\sigma_z$ are the respective Pauli spin matrices. We
define $V(y)$ as above, rather than the more symmetrical $V(y) = \sigma_y$
(which would correspond to a different factor system within the same cohomology
class), because the factor system of \eqnref{D2proj} will turn out to be closely
connected to the conventional formulation of the KT transformation.

\section{The generalized Kennedy-Tasaki transformation}
\label{sec_generalized_kt}
In this section, we will define our generalized KT transformation,
for an SPT phase characterized by a cohomology class $[\omega]$ and an on-site
symmetry representation of a group $G$. We will require that $G$ be finite and
abelian, and that the cohomology class $[\omega]$ be maximally non-commutative
(to be defined below).
A special property of
an abelian symmetry is that the irreps are one-dimensional; therefore, the
on-site representation $u(g)$ must decompose as
\begin{equation}
\label{eq:irreps}
u(g) = \bigoplus_{\chi} \chi(g) \mathbb{I}_{m_{\chi}}
\end{equation}
where the sum is over the one-dimensional representations (characters) $\chi$ of $G$. For
simplicity of presentation, we assume that none of the multiplicities $m_{\chi}$ are greater than
1; thus, we can write
\begin{equation}
u(g) = \sum_{\chi} \chi(g) \ket{\chi} \bra{\chi}
\end{equation}
where the $\{ \ket{\chi} \}$ form an orthonormal basis, and the sum is over
those $\chi$ such that $m_{\chi} > 0$. However, all the results of this paper can easily
be generalized to the case of multiplicities greater than 1.  For the Haldane phase,  this basis $\{ \ket{\chi} \}$ is the basis $\{ \ket{x}, \ket{y}, \ket{z} \}$ discussed in Sec.~\ref{sec:KTtrans}.

Our generalized construction applies for any SPT phase with respect to the
aforementioned symmetry, so long as the corresponding cohomology class $[\omega]$
is \emph{maximally-noncommutative}, which is to say that the subgroup $G(\omega)
= \{ g \in G : V_\omega(g) V_\omega(h) = V_\omega(h) V_\omega(g) \ \forall\ h \in G \}$ is trivial.
This property does not depend on the choice of
representative factor system for the cohomology class.
(Throughout this section, we will use $V_\omega$ to denote some projective
representation of $G$ with factor system $\omega$; 
it does not matter how the projective
representation is chosen because we only use the multiplicative relations between the
matrices $V_\omega(g)$, and these are determined by $\omega$.) As follows from Refs.~\onlinecite{frucht1932,berkovich_characters}, a finite abelian group $G$ will have at least one maximally non-commutative factor system if and only if it is of ``symmetric type'', i.e. $G \cong H \times H$
for some group $H$. Of course, even if the full symmetry group is not of this form, then it might still have a subgroup of symmetric type, for which our
method could be applied.

An important property of a maximally non-commutative factor system is the
following. Any cohomology class for an abelian group can be considered to induce a
homomorphism $\varphi_\omega$
from $G$ to $G^{*}$ (where $G^{*}$ is the
\emph{character group} of $G$, i.e.\ the group of one-dimensional projective
representations of $G$ under multiplication), according to 
\begin{equation}
\label{varphi_defn}
\varphi_\omega(g) = \chi_g^{\omega},
\end{equation}
where $\chi_g^{\omega}$ is the one-dimensional representation of $G$ such that
\begin{equation}
\label{chig}
V_\omega(g^{\prime}) V_\omega(g) V_\omega(g^{\prime})^{\dagger} =
\chi_g^\omega(g^{\prime}) V_\omega(g).
\end{equation}
Observe that from \eqnref{chig} one can prove both that $\chi_g^\omega(g_1^{\prime})
\chi_g^\omega(g_2^{\prime}) = \chi_g^\omega(g_1^{\prime} g_2^{\prime})$ (i.e.\ 
$\chi_g^\omega = \varphi_\omega(g)$ is in $G^{*}$) and that $\chi_{g_1}^\omega
\chi_{g_2}^\omega = \chi_{g_1 g_2}^\omega$ (i.e.\ $\varphi_\omega$ is a
homomorphism).
For the particular case of a maximally non-commutative projective representation
of a finite abelian group,
the kernel of $\varphi_\omega$ [which is equal to $G(\omega)$ in
general] is trivial, and therefore $\varphi_\omega$ is an \emph{iso}morphism; that is, for any $\chi \in
G^{*}$ there is a unique $g \in G$ such that $\chi_g^{\omega} = \chi$.

We construct the unitary $\mathcal{D}_\omega$ corresponding to the generalized
Kennedy-Tasaki transformation (as we will see later, it maps \emph{from} the maximal
symmetry-breaking phase into the SPT phase), acting on a chain of $N$ sites with open boundary
conditions, according to
\begin{equation}
\label{generalized_KT}
\mathcal{D}_{\omega} = \sum_{\mathbf{\chi}} \Omega_\omega\bigl(\varphi_\omega^{-1}(\bchi)\bigr) \ket{\bchi}
\bra{\bchi},
\end{equation}
where we use the abbreviations $\bchi = (\chi_1, \ldots, \chi_N)$,
$\varphi_\omega^{-1}(\bchi) = \bigl(\varphi_\omega^{-1}(\chi_1), \ldots,
\varphi_\omega^{-1}(\chi_N)\bigr)$, and $\Omega_\omega(\mathbf{g})$
is the phase factor defined such that
\begin{equation}
\label{Omega_defn}
V_\omega(g_N) \cdots V_\omega(g_1) = \Omega_\omega\bigl(\mathbf{g}) V_\omega(g_N \cdots g_1\bigr).
\end{equation}
(here, as in \eqnref{eq:irreps}, the sum is over the characters $\chi$ that appear in
the representation.)

For the case of a spin-1 chain with $D_2$ symmetry, one can check directly that the choice of factor system $\omega$ defined by the projective representation \eqnref{D2proj} gives
\begin{equation}
  \label{eq:D2coho}
  V(a_N)\cdots V(a_1) = (-1)^{\sum_{j < k} \mu(a_j) \nu(a_k)} V(a_N \cdots a_1) ,
\end{equation}
(where, loosely, one obtains a phase factor of $-1$ for every $V(z) = \sigma_z$ operator to the left of a $V(x) = \sigma_x$ operator).  Therefore, \eqnref{generalized_KT}
reduces to the standard Kennedy-Tasaki transformation \eqnref{DKT_basis} if we choose this
factor system.
Note that the definition of $\mathcal{D}_{\omega}$ is not the same for
different factor systems $\omega$ within the same cohomology class. However, the
difference is not very significant; see Appendix \ref{choice_of_factor_system}.

Due to the way the unitary $\mathcal{D}_\omega$ is defined, we can immediately derive the basic property that, although it is
a non-local transformation,
for any
\emph{symmetry-respecting} observable $A$ supported on a block of $n$ sites, the
transformed observable $\mathcal{D} A \mathcal{D}^{\dagger}$ is still supported
on the same block.
We will use the notation $\bchi = (\bchi_l, \bchi_b,
\bchi_r)$, corresponding to grouping the sites in the chain according to
whether they are, respectively, to the left of, within, or to the right
of the block containing the support of $A$. Thus the matrix element $
\bra{\bchi_l, \bchi_b, \bchi_r} A \ket{\bxi_l, \bxi_b,
\bxi_r}$ can be nonzero only if $\bchi_l = \bxi_l$,
$\bchi_r = \bxi_r$, and $\prod_{j=1}^n \chi_{b,j} = \prod_{j=1}^n
\xi_{b,j}$ (the last condition comes from the assumption that $A$ commutes with
the symmetry). As a result, it is easy to show from the definition of $\Omega$
[\eqnref{Omega_defn}] that $\Omega_\omega\bigl(\varphi^{-1}_\omega(\bchi)\bigr) \
\Omega_\omega\bigl(\varphi_\omega^{-1}(\bxi)\bigr)^{-1} = 
\Omega_\omega^{(n)}\bigl(\varphi_\omega^{-1}(\bchi_b)\bigr)
\Omega_\omega^{(n)}\bigl(\varphi_\omega^{-1}(\bxi_b)\bigr)^{-1}$, and hence that
$\mathcal{D}_\omega A \mathcal{D}_\omega^{\dagger} =
\mathcal{D}_\omega^{(n)} A \mathcal{D}_\omega^{(n){\dagger}}$,
 where
$\Omega_\omega^{(n)}$ and $\mathcal{D}_\omega^{(n)}$ are defined as
$\Omega_\omega$ and $\mathcal{D}_\omega$ would be if the $n$ sites in the
block constituted the entire chain.

\section{Action of the generalized KT transformation on a generalized AKLT state}
\label{generalized_aklt_action}
Although we have focussed on the transformation of the Hamiltonian
under the generalized KT transformation $\mathcal{D}_{\omega}$, for illustrative purposes we will consider in this section a particular Hamiltonian within the SPT phase described by cohomology
class $[\omega]$, for which the ground state subspace can be found analytically.  We calculate explicitly how this subspace transforms under
$\mathcal{D}_{\omega}^{\dagger}$, and show that the transformed ground state
subspace reflects the spontaneous breaking of the symmetry in the bulk. The
definition of $\mathcal{D}_{\omega}$ [Eqs.\ (\ref{generalized_KT}) and
(\ref{Omega_defn})] arises naturally out of this discussion.

Affleck, Kennedy, Lieb and Tasaki (AKLT) \cite{aklt1,*aklt2} constructed a system in the
Haldane phase for which the
ground state can be represented exactly as a ``valence-bond solid'', or (in more
modern language) a \emph{matrix-product state} (MPS) \cite{perezgarcia_mps}. We will now define a generalization of the
AKLT ground state for the SPT phase with cohomology class $[\omega]$. We will
write it for open boundary conditions, which means we have to define a
subspace $\mathcal{P}$ of states corresponding to the degenerate ground-state subspace. The states in this subspace are of the MPS form
\begin{equation}
  \label{eq:generalized_aklt}
\sum_{\bchi} \Tr(A_{\chi_N} \cdots A_{\chi_1} B) \ket{\bchi},
\end{equation}
where we set $A_{\chi} = V_\omega\bigl(\varphi_\omega^{-1}(\chi)\bigr)$ (where
$V_\omega$ is an irreducible projective representation with factor system
$\omega$), and the subspace comprises the states obtained from all possible $D \times D$ matrices $B$ (with $D$ the dimension of
$V_\omega$). The theory of MPS parent Hamiltonians \cite{perezgarcia_mps} allows one to
construct a local frustration-free Hamiltonian for which $\mathcal{P}$ is the $D^2$-fold degenerate ground state subspace.
From the classification of SPT order in matrix-product states
\cite{schuch,chen_gu_wen}, one can show
\cite{else_schwarz_bartlett_doherty_symmetry} that $H$ indeed lies in the SPT
phase described by cohomology class $[\omega]$.

We will only consider the state $\ket{\Psi}$ resulting from setting $B = \frac{1}{D}\sum_g
V(g)^{\dagger}$, as it turns out that applying $[u(g)]^{\otimes N}$ to
$\ket{\Psi}$ for group elements $g$ generates a basis for $\mathcal{P}$. This gives
\begin{equation}
\ket{\Psi} = \sum_{\bchi} \Omega_\omega\bigl(\varphi_\omega^{-1}(\bchi)\bigr)
\ket{\bchi},
\end{equation}
where we have used the fact that $\Tr[ V_\omega(h) V_\omega(g)^{\dagger}] = D
\delta_{g,h}$ (which follows from the fact that $\Tr V_\omega(g) = 0$ for $g \neq 1$, a consequence of maximal non-commutativity). This means that
\begin{equation}
\mathcal{D}_\omega^{\dagger} \ket{\Psi} = \sum_{\bchi} \ket{\bchi} = \ket{\phi}^{\otimes N},
\end{equation}
where $\ket{\phi} = \sum_{\chi} \ket{\chi}$. Since $[u(g)]^{\otimes N}$ commutes
with $\mathcal{D}_{\omega}^{\dagger}$, a basis for the transformed subspace
$\mathcal{D}_\omega^{\dagger} \mathcal{P}$ comprises the states $\{
[u(g)]^{\otimes N} \ket{\phi}^{\otimes N}, g \in G \}$. Thus the transformed
Hamiltonian under $\mathcal{D}_{\omega}^{\dagger}$ indeed has a set of locally
distinguishable symmetry-breaking ground states, as we expect.

\section{String order}
\label{sec_string_order}
A key property of the Kennedy-Tasaki transformation is that it relates
two-particle correlations (which are expected to be long-ranged in the
maximal symmetry-breaking phase for the $Z_2 \times Z_2$ symmetry) to the string correlation functions that characterize 
the Haldane phase. Here, we will establish a similar correspondence for our
general construction. This property will also allow us to determine how the generalized
Kennedy-Tasaki transformation maps between different quantum phases.

\subsection{Symmetry-breaking phases and two-particle correlations}
\label{subsec_twoparticle}
Let us first give a general discussion of the two-particle correlations that we expect to
see in the maximal symmetry-breaking phase (i.e.\ where the subgroup of symmetry
operations that are \emph{not} spontaneously broken in the bulk is trivial) for
an on-site abelian symmetry.
A system in this phase will have a collection of degenerate symmetry-breaking
ground states. Traditionally, the symmetry-breaking is detected through the
nonzero value of an \emph{order parameter}, which is the expectation value of a
single-site observable $A$ such that $\langle A \rangle = 0$ for any symmetry-\emph{respecting} state.
For example, in the case of the quantum
transverse-field Ising model on spin-$1/2$'s, with Hamiltonian
\begin{equation}
H = -\sum_{i} \sigma_i^z \sigma_{i+1}^z + \lambda \sum_i \sigma_i^x,
\end{equation}
the appropriate order parameter is $\langle \sigma_z \rangle$.  The fact that
$\sigma_x \sigma_z \sigma_x = -\sigma_z$  ensures that $\langle \sigma_z \rangle$
must be zero for any state respecting the spin-flip symmetry $\prod_i
\sigma_i^x$. In the general case, we can consider an observable $A$ such that
\begin{equation}
\label{Achi_defn}
  u(g^{\prime}) A u(g^{\prime})^{\dagger} = \chi(g^{\prime}) A
\end{equation}
for some $\chi \in G^{*}$.  By a similar argument as before, for $\chi \neq 1$ we find that
$\langle A \rangle = 0$ for any symmetry-respecting state. As another example,
in the case of a spin-1 chain with the $D_2$ rotation symmetry,
the spin-component operators $S_a$ ($a = x,y,z$) satisfy \eqnref{Achi_defn} for
appropriate choices of $\chi$.

Denote the space of operators $A$ satisfying \eqnref{Achi_defn} by $\mathcal{A}_\chi$.
For a given symmetry-breaking state, a given operator in $\mathcal{A}_{\chi}$
could still have zero expectation value by accident. However, we will now argue that, 
for a given maximal symmetry-breaking state, for
\emph{every} non-trivial $\chi \in G^{*}$ a generic choice of
$A \in \mathcal{A}_\chi$ will reveal the symmetry-breaking through its
nonzero expectation value. Indeed, let $\rho$ be the reduced state density
operator on a single site. It suffices to show that the subspace $\mathcal{B}_{\chi} = \{ A \in \mathcal{A}_{\chi} | \mathrm{Tr}(A \rho) = 0 \}$ is a proper subspace (i.e.\ $\mathcal{B}_{\chi} \neq \mathcal{A}_{\chi}$). Suppose by way of contradiction that $\mathcal{B}_\chi = \mathcal{A}_\chi$. Then \emph{every} $A \in \mathcal{A}_{\chi}$ must satisfy $\mathrm{Tr}(A \rho) = 0$. But since the set $\{
\ket{\chi^{\prime}} \bra{\chi^{\prime} \chi} : \chi^{\prime} \in G^{*} \}$ comprises a basis
for $\mathcal{A}_\chi$, this would imply that
$\bra{\chi^{\prime} \chi} \rho \ket{\chi^{\prime}} = 0$ for all
$\chi^{\prime}$. In the maximal symmetry breaking phase, the ground state has no
residual symmetry, and hence there is no constraint on the reduced state $\rho$ that would force
all of these matrix elements to be zero [whereas if the ground state were
invariant under the symmetry operation corresponding to the group element $g$,
then this would force $\bra{\chi_1} \rho \ket{\chi_2} = 0$ for all
$\chi_1,\chi_2 \in G^{*}$ such that $\chi_1(g) \neq \chi_2(g)$.] Certainly, for
a \emph{generic} state in the maximal
symmetry-breaking phase, these matrix elements would not all be zero.

An alternative
measure of the symmetry-breaking is the two-particle
correlation function $\langle C_n(A,B) \rangle$ (for $A,B \in
\mathcal{A}_\chi$), where
\begin{equation}
\label{CnAB}
C_n(A,B) \equiv A^{\dagger} \otimes \mathbb{I}^{\otimes (n-2)} \otimes B.
\end{equation}
Because each of the symmetry-breaking ground states should be short-range
correlated, the correlation function $\langle C_n(A,B) \rangle$ converges
to $\langle A \rangle^{*} \langle B \rangle$ as $n \to \infty$. The expectations here
are taken with respect to a particular choice of symmetry-breaking ground state, but
notice that $C_n(A,B)$ commutes with the symmetry, and therefore its
expectation is independent of this choice.

\subsection{SPT phases and string correlation functions}
A key feature of the Haldane phase and its generalizations is that there is no locally-detectable symmetry breaking in the bulk, and consequently all the two-particle correlations $\langle C_n(A,B) \rangle$ decay exponentially as $n \to \infty$. Nevertheless, such SPT phases still have a more subtle form of long-range order detectable through \emph{string} correlations. As a result, we are led to consider the following generalization of \eqnref{CnAB} (reducing to it when $g = 1$):
\begin{equation}
\label{CnABg}
C_n(A,B;g) = A^{\dagger} \otimes [u(g)]^{\otimes (n-2)} \otimes B.
\end{equation}
(Recall that $u(g)$ is the unitary on-site action of the symmetry.)
In particular, the den Nijs-Rommelse string operators \cite{aklt_string_order} for the Haldane phase,
\begin{equation}
S^\alpha \otimes [e^{i \pi S^\alpha}]^{\otimes (n-2)} \otimes S^\alpha = C_n(S^\alpha, S^\alpha; \alpha), \quad \alpha = x,y,z,
\end{equation}
with $S^\alpha$ the spin-component operators, are of this form.
In the Haldane phase, the den Nijs-Rommelse string correlations are long-ranged, i.e.\ $\lim_{n \to \infty} 
\langle C_n(S^\alpha, S^\alpha; \alpha) \rangle \neq 0$.

Traditionally, long-ranged string correlations have been viewed as evidence of
non-trivial order. However, there is a need for caution: the limiting string
correlation functions $\lim_{n \to \infty} \langle C_n(A,B;g) \rangle$ are
nonzero for generic choices of $A$ and $B$ whenever the symmetry is unbroken in
the bulk, and need not reflect any non-trivial SPT order
\cite{string_order_symmetries,detection_string_order}. Therefore, in order to
obtain useful criteria for identifying SPT phases, we must restrict ourselves to
restricted classes of $A$ and $B$. Indeed, it turns out to be useful to require,
as in the symmetry-breaking case, $A,B \in \mathcal{A}_{\chi}$ for some linear
character $\chi$. In that case, the selection rule discussed in
Ref.~\onlinecite{detection_string_order} forces $\lim_{n \to \infty} \langle C_n(A,B;g)
\rangle = 0$ when $\varphi_\omega(g) \neq \chi$, where ${\varphi}_\omega$ is the
homomorphism induced by the cohomology class [\eqnref{varphi_defn}] (see
Appendix \ref{appendix_pattern_of_string_order} for the proof). In the case
$\varphi_\omega(g) = \chi$, there is no such selection rule and so we expect
that the corresponding string correlation will generically be long-ranged. (It
can be checked that the latter case is the relevant one for the den
Nijs-Rommelse string correlations in the Haldane phase.) Thus, the pattern of
long-ranged string orders of the form considered is a useful way of identifying
phases; we will make this idea more precise in Section \ref{subsec_signature}.

\subsection{Mapping of correlation functions under the generalized Kennedy-Tasaki transformation}\label{subsec:mapping_correlation}
We have established that the string operators $C_n(A,B;g)$ are useful tools for
identifying phases. Therefore, it makes sense to calculate how these operators
transform under the generalized KT transformation $\mathcal{D}_\omega$, where
$[\omega]$ is a maximally non-commutative cohomology class. This calculation is
done in Appendix \ref{proof_of_transformed_string}; the result is (for $A,B \in \mathcal{A}_\chi$)
\begin{equation}
\label{transformed_string}
\mathcal{D}_\omega C_n(A,B;g) \mathcal{D}_\omega^{\dagger} = C_n(\widetilde{A},
\widetilde{B}; \widetilde{g})
\end{equation}
with
\begin{equation}
\label{transformed_string_values}
\widetilde{A} = A W_\omega(\chi)^{\dagger}, \quad \widetilde{B}  = 
B W_\omega^{\prime}(\chi)^{\dagger}, \quad \widetilde{g} = g \varphi_\omega^{-1}(\chi),
\end{equation}
where 
\begin{align}
  W_\omega(\chi) &= \sum_{\chi'} \omega\bigl(\varphi_\omega^{-1}(\chi),
\varphi_\omega^{-1}(\chi^{\prime}) \bigr)  \ket{\chi^{\prime}} \bra{\chi^{\prime}} , \\
W_\omega^{\prime}(\chi) &= \sum_{\chi'} \omega\bigl(\varphi_\omega^{-1}(\chi^\prime),
\varphi_\omega^{-1}(\chi) \bigr)  \ket{\chi^{\prime}} \bra{\chi^{\prime}} ,
\end{align}
(observe that $A,B \in \mathcal{A}_\chi$ implies $\widetilde{A},\widetilde{B}
\in \mathcal{A}_{\chi}$ as well, since $W_\omega$ and $W_\omega^{\prime}$ commute with the symmetry).

To see the significance of this result, suppose that $[\omega]$ is a maximally non-commutative cohomology class. Recall that, in a system with maximal symmetry breaking, for any $\chi \neq 1$ we expect to be able to find $A,B \in \mathcal{A}_\chi$ such that $\lim_{n \to \infty} 
\langle C_n(A,B;1) \rangle \neq 0$. But then, by \eqnref{transformed_string}, this implies that in the transformed system obtained from the original one by $\mathcal{D}_\omega$, we will have $\lim_{n \to \infty} \langle C_n(\widetilde{A}, \widetilde{B}; \varphi_\omega^{-1}(\chi)) \rangle \neq 0$, i.e.\ there are long-ranged string correlations of precisely the form that we expect to get in the SPT phase characterized by cohomology class $[\omega]$. In the next subsection, we will turn this into a proof that the transformed system is indeed in that phase.

\subsection{Patterns of string order as a ``signature'' for quantum phases}
\label{subsec_signature}
We have already seen that the long-range behavior of string correlations of the
form $\langle C(A,B;g) \rangle$ (with $A,B \in \mathcal{A}_\chi$ for some
character $\chi$) is a useful probe for identifying different kinds of ordering
in systems with a finite abelian on-site symmetry. In Appendix
\ref{appendix_pattern_of_string_order}, we go
further, and show that this long-range behavior \emph{uniquely} identifies all
possible quantum phases that result from symmetry-respecting Hamiltonians. (The
general classification of such phases was given in Ref.~\onlinecite{chen_gu_wen_2,schuch}; it includes conventional symmetry-breaking phases, SPT phases with no symmetry-breaking in the bulk, as well as other examples in which SPT and symmetry-breaking orders combine.)

The result of Appendix~\ref{appendix_pattern_of_string_order} is expressed in terms of the following ``signature'' function $M$ acting on $G^{*} \times G$ to measure which of the string correlations are long-ranged:
\begin{equation}
\label{Mchig}
M(\chi,g) = \begin{cases} 1 & \text{if $\lim_{n \to \infty} \langle C_n(A,B,g)
\rangle \neq 0$
\emph{generically}} \\
{} & \text{\quad in the phase, when $A,B \in \mathcal{A}_\chi$}, \\
0 & \text{otherwise}
\end{cases}
\end{equation}
We have included the word ``generically'', because it is possible that there
might be specific points in the phase and/or choices of $A,B \in
\mathcal{A}_{\chi}$ such that the limiting correlation is ``accidentally'' zero.
[For example, in the case $g = 1$, we derived in
\secref{subsec_twoparticle} the condition for a given state in the maximal
symmetry-breaking phase to satisfy $\lim_{n \to \infty} C_n(A,B;1) \rangle = 0$
for all $A,B \in \mathcal{A}_{\chi}$, even though generically we expect these
two-body correlations to be long-ranged.] The result of Appendix~\ref{appendix_pattern_of_string_order} is then that
each possible phase in the general classification has a distinct signature $M$.

Combining this result with that of the previous subsection~\ref{subsec:mapping_correlation} allows us to definitively establish in general how different phases are transformed into each other by the 
generalized KT transformation $\mathcal{D}_{\omega}$ (with $[\omega]$ a maximally non-commutative cohomology class). Indeed, suppose we start from a phase
described by signature $M$. Then, by Eqs.~(\ref{transformed_string}) and (\ref{transformed_string_values}), the transformed phase resulting from application of $\mathcal{D}_{\omega}$ has signature 
\begin{equation}
\label{signature_mapping}
M^{\prime}(\chi,g) = M(\chi, g [\varphi_\omega^{-1}(\chi)]^{-1}).
\end{equation}

In particular, we can consider the case that the starting phase is the maximal
symmetry breaking phase (all the symmetries broken in the bulk). The arguments
of \secref{subsec_twoparticle} show that for such a phase, $M(\chi,1) = 1$ for
all $\chi$. Furthermore, it was shown in Ref.~\onlinecite{detection_string_order} that
string order of the form $\langle C_n(A,B;g) \rangle$ can be long-ranged only
when the symmetry corresponding to $g$ is unbroken in the bulk. Thus, the
maximal symmetry-breaking phase has signature $M(\chi,g) = 1 \Leftrightarrow g =
1$. It follows that the transformed phase resulting from applying
$\mathcal{D}_\omega$ has signature $M^{\prime}(\chi,g) = 1 \Leftrightarrow g =
\varphi_\omega^{-1}(\chi)$. From the discussion of Appendix
\ref{appendix_pattern_of_string_order}, we see that
this is precisely the signature of the SPT phase with cohomology class
$[\omega]$, as expected.

Note that, although in this paper we have concentrated on the duality between pure SPT order and maximal symmetry-breaking order, \eqnref{signature_mapping} can be used to determine in general how the generalized KT transformation relates $\mathcal{D}_\omega$ different symmetric phases to each other, including combined symmetry-breaking/SPT phases. For example, see Ref.~\cite{duivenvoorden_quella} for a discussion of the $Z_N \times Z_N$ case. (In this case, our generalized KT transformation reduces to the one defined in Ref.~\cite{duivenvoorden_quella}, or a variant thereof, depending on which cohomology class $[\omega]$ and factor system representative $\omega$ one uses in the construction).

\section{The Kennedy-Tasaki transformation for continuous symmetries}
\label{sec:examples}
We stress that our assumption of a maximally-noncommutative
cohomology class of a finite abelian group might not be as restrictive as it
sounds. Indeed, an SPT phase characterized by an arbitrary group could still be
identified as part of a maximally non-commutative SPT phase with respect to a finite abelian subgroup.
As an example, here we will discuss how our framework allows us to apply the concept of hidden
symmetry breaking to some generalizations of the Haldane phase. 

Just as the
Haldane phase is motivated by $\SO(3)$-invariant antiferromagnets, these
generalized Haldane phases contain systems that are invariant under an
$\SO(2k+1)$ or $\SU(k)$ symmetry.
However, in each case, we will identify a finite abelian subgroup (analogous to
$D_2$ for the Haldane phase), which will turn out to be the relevant one for
identifying the hidden symmetry breaking. In each case, this finite abelian
subgroup will turn out to be sufficient to classify the phases, since imposing
the full continuous symmetry does not separate any phases that could not
already be distinguished through this subgroup. This suggests that, even when
the full continuous symmetry is present, we should describe the SPT order in
terms of the hidden breaking of the finite abelian subgroup.

\subsection{The $\SO(2k+1)$ Haldane phase}
\label{SO2kp1}
For systems invariant under an on-site $\SO(2k+1)$ symmetry, there is exactly one
non-trivial SPT phase \cite{SO_string_order,suN_phases}, which we can think of as a generalization of the
Haldane phase (reducing to it in the case $k = 1$). The corresponding cohomology
class is that of the spinor representations of $\SO(2k+1)$ (which are, in fact,
\emph{projective} representations). 


Identifying $\SO(2k+1)$ with its representation in terms of $(2k+1) \times
(2k+1)$ orthogonal matrices with unit determinant, we define $G_k \equiv \{ A \in \SO(2k+1) : A
\text{ is diagonal in the standard basis} \}$, which constitutes
a finite abelian subgroup. We can construct a minimal
set of generators $\{ u^{(l)}, l = 1,\ldots,2k \}$ with matrix elements 
\begin{equation}
u^{(l)}_{i,j} = (-1)^{1-\delta_{i,l}} \delta_{i,j}
\end{equation}
(we do not include $u^{(2k+1)}$ in our of minimal set of generators because it is not
independent of the rest; indeed, $u^{(2k+1)} = \prod_{l=1}^{2k} u^{(l)}$). This
shows that $G_k \cong Z_2^{\times 2k}$. It can be shown, by considering the
restriction of the spinor representations of $\SO(2k+1)$ to the subgroup $G_k$, that the cohomology
class of $G_k$ for systems in the non-trivial SPT phase with respect to
$\SO(2k+1)$ is that of the projective representation generated by
\begin{equation}
V(u^{(l)}) = \Gamma_l,
\end{equation}
where the $2k$ matrices $\Gamma_l$ obey the anti-commutation relations $\{
\Gamma_a, \Gamma_b \} = 2 \delta_{a,b}$. It is straightforward to show that this
cohomology class is maximally non-commutative.

Thus, we can use our general prescription [\eqnref{generalized_KT}] to construct
a generalized KT transformation for systems in the non-trivial SPT phase with
respect to $\SO(2k+1)$. In analogy to the original KT transformation for the
Haldane phase (which breaks the full rotation symmetry, preserving only the
discrete subgroup $D_2$), the resulting transformed system will only have the
discrete $Z_2^{\times 2k}$ symmetry instead of the full $\SO(2k+1)$.
Furthermore, the transformed system will be in a maximal local symmetry-breaking
phase for this discrete symmetry. In this sense, the non-trivial $\SO(2k+1)$ SPT
phase can be understood as a result of the ``hidden breaking'' of the $Z_2^{\times
2k}$ symmetry. Note that, if we make a particular choice of factor system within the appropriate cohomology class,
it can be shown that the generalized Kennedy-Tasaki transformation constructed
according to our general prescription [see~\eqnref{generalized_KT}] coincides with
the one constructed in Ref.~\onlinecite{SO_string_order}.

We remark that, since for the $\SO(2k+1)$ symmetry group there is only one nontrivial SPT phase, it can already be distinguished from the trivial phase via a $Z_2 \times Z_2$ subgroup. Although one could therefore construct the generalized KT transformation $\mathcal{D}_\omega$ based on the $Z_2 \times Z_2$ subgroup, as in Ref. \onlinecite{duivenvoorden_quella}, we prefer to construct it based on $Z_2^{\times 2k}$. This ensures that the transformed phase is maximally symmetry-breaking. If one instead uses only the $Z_2 \times Z_2$ subgroup to construct $\mathcal{D}_\omega$, it can be shown (using similar arguments to \secref{subsec_signature} and Appendices \ref{proof_of_transformed_string} and \ref{appendix_pattern_of_string_order}) that the resulting phase breaks $Z_2 \times Z_2$, but is still SPT-ordered with respect to the remaining $Z_2^{\times 2(k-1)}$.

In general, a useful way to ensure
that the transformed phase has no residual SPT order is by counting the ground state
degeneracy: if the degeneracy of the original SPT phase is fully explained by
the symmetry-breaking in the transformed phase, then there cannot be any
residual SPT order. For instance, consider the $\SO(2k+1)$ case. We can assume
that the fractionalized representation of the symmetry on the edge is the fundamental
spinor representation, as it
suffices to confirm the lack of residual SPT order at a single point in the
phase. There is then a $2^k$-fold degeneracy associated with
each edge, which agrees with the $2^{2k}$-fold degeneracy we expect for a phase that
maximally breaks a $Z_2^{\times 2k}$ symmetry. A similar property also holds for
the $\SU(k)$ example considered in the next section.

\subsection{SPT phases for $\SU(k)$}
Instead of thinking of the Haldane phase as invariant under an $\SO(3)$ symmetry,
we can also think of it as invariant under $\PSU(2) \equiv \SU(2)/\{ +1, -1 \}$. Of
course, $\PSU(2) \cong \SO(3)$, but this suggests an alternative generalization
of the Haldane phase: one that is invariant under an on-site representation of 
$\PSU(k) \equiv \SU(k)/C_k$, where $C_k = \{ \exp(2i\pi l/k) : l =
0,\ldots,k-1\}$.

Given the definition of $\PSU(k)$, we can construct a finite
abelian subgroup by identifying a subgroup of $\SU(k)$ that is abelian up to
phase factors (i.e.\ up to elements of $C_k$). The discrete Heisenberg-Weyl group is
such a subgroup; it is the group generated [in the standard representation of
$\SU(k)$] by the two operators
\begin{align}
X &= \frac{1}{W} \sum_{l=0}^{k-1} \ket{(l+1) \operatorname{mod} k} \bra{l}, \\ 
Z &= \frac{1}{W} \sum_{l=0}^{k-1} w^l \ket{l} \bra{l}
\end{align}
where $w$ is a primitive $k$-th root of unity, and we have included the normalization factor $W =
w^{(k-1)/2}$ to ensure that $\det X = \det Z = 1$. The fact that the subgroup generated by
$X$ and $Z$ is abelian up to phases follows from the relation $ZX = w XZ$. The
abelian subgroup of $\PSU(k)$ corresponding to the Heisenberg-Weyl
group is isomorphic to $Z_k \times Z_k$. The cohomology group for $Z_k
\times Z_k$ is $H^2(Z_k \times Z_k, \U(1)) \cong Z_k \cong H^2(\PSU(k), \U(1))$,
and it can be shown\cite{duivenvoorden_quella}
that the $k$ cohomology classes of $\PSU(k)$ correspond exactly to the $k$
cohomology classes of the $Z_k \times Z_k$ subgroup \cite{}.
Thus, the $Z_k \times Z_k$ subgroup is sufficient
to characterize all the SPT phases even in the presence of the full $\PSU(k)$ symmetry. In order to apply our generalized
Kennedy-Tasaki transformation, we need a maximally non-commutative cohomology
class; if we let $[\omega_0]$ be a generator for the cohomology
group $H^2(Z_k \times Z_k, \U(1))$, then it can be shown that $[\omega_0^l]$ is
maximally non-commutative if and only if $l$ and $k$ are coprime.

\subsection{Other continuous symmetry groups}

We leave it as an
open question whether a similar analysis to the above for $SO(2k+1)$ and $SU(k)$ holds for other continuous symmetry groups. In Ref.~\onlinecite{duivenvoorden_quella}, it is shown that a subgroup of the form $Z_N \times Z_N$ can be found for all the cases involving classical Lie groups.  However, as we have shown with the $SO(2k+1)$ example in~\secref{SO2kp1}, this is not the whole story, especially if the aim is to identify a relevant generalised KT transformation that can `fully' remove the SPT order, i.e., that can relate a SPT-ordered phase to a maximally symmetry-breaking phase.

\section{The topological disentangler}
\begin{figure}
\includegraphics[width=8.5cm]{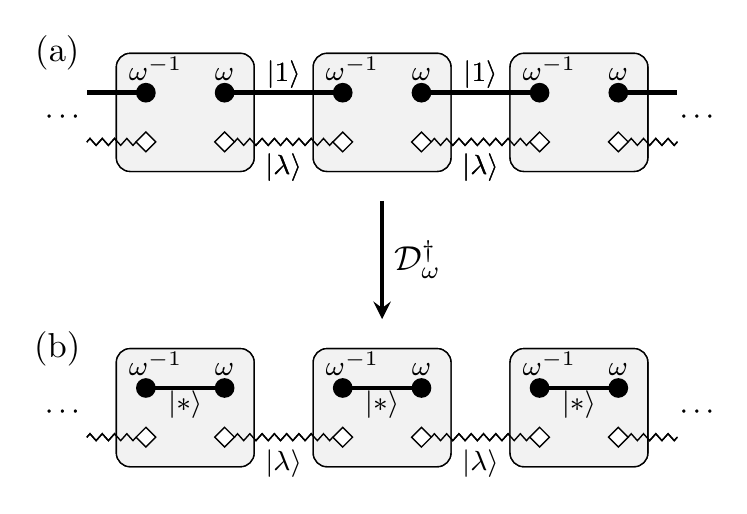}
\caption{\label{fig:topological_disentangler} (a) The ``dimer state'' renormalization fixed point
for the SPT phase corresponding to a maximally commutative cohomology class
$[\omega]$; (b) The result of applying $\mathcal{D}_\omega^{\dagger}$, for a particular
choice of boundary conditions. Each shaded area represents one coarse-grained site. The black dots transform under irreducible projective
representations with factor systems $\omega$ and $\omega^{-1}$ under the symmetry, and the
diamonds do not transform at all under the symmetry. Note: Two adjacent black dots
transform linearly under the symmetry; therefore, we can introduce the
simultaneous eigenbasis $\{ \ket{ \chi} \}$ of the symmetry (with the states
labelled by linear characters $\chi$; from Schur's Lemma it follows that they
must be maximally entangled). For $\chi = 1$ this gives the state
$\ket{1}$ appearing in (a). In (b), we have defined $\ket{*} = \sum_\chi
\ket{\chi}$. The state $\ket{\lambda}$ is not universal and depends on the specific point in the phase.}
\end{figure}

In this section, we will briefly outline a physical
interpretation of the resulting correspondence between the SPT ground states and
the symmetry-breaking ground states, in terms of the entanglement structure of the ground states.

When we group sites together in blocks of size $\gg \xi$, with $\xi$ the
correlation length, any gapped ground state starts to resemble (up to on-site
unitary rotations on the blocked sites) a ``dimer state'' which can be viewed
as a renormalization fixed point~\cite{verstraete}. If the ground state is in an SPT phase
characterized by a maximally non-commutative cohomology class, this dimer state
will take the form shown in \figref{fig:topological_disentangler}(a); this is a consequence of the fact that a maximally non-commutative factor system corresponds to a unique irreducible
projective representation~\cite{frucht1932,berkovich_characters}. We see that the entanglement between two halves of the chain has two origins:
the universal ``topological'' entanglement represented by the maximally entangled state
$\ket{1}$, and the ``non-topological'' entanglement represented by the state
$\ket{\lambda}$. (The ``dot'' and ``diamond'' particles correspond to the
``protected'' and ``junk'' subsystems discussed in
Ref.~\onlinecite{else_schwarz_bartlett_doherty_symmetry}.)

It is therefore instructive to consider what happens under
$\mathcal{D}_\omega^\dagger$ to a ground state of the form shown in
\figref{fig:topological_disentangler}(a). The calculation required is similar to that of
\secref{generalized_aklt_action}. (Indeed, when written in a matrix-product state form, the state of
\figref{fig:topological_disentangler}(a) reduces to \eqnref{eq:generalized_aklt} when the ``extra'' particles in the
$\ket{\lambda}$ state are absent.) Here we
just state the result: for a suitable choice of boundary conditions, the
resulting state is as depicted in \figref{fig:topological_disentangler}(b); note that this resulting state is no longer
invariant under the symmetry, and the orbit of this state under the symmetry is the set of
symmetry-breaking ground states for the transformed system. We see that the topological
component of the entanglement has been eliminated, with the non-topological part
of the state remaining untouched. In this sense, we
can think of the generalized KT transformation $\mathcal{D}_{\omega}^{\dagger}$
as a ``topological disentangler'' \cite{topological_disentangler}.

\section{Discussion}
We have presented a generalization of the Kennedy-Tasaki transformation, which maps certain one-dimensional models with SPT order to ones with traditional symmetry-breaking.  This formulation further expands the characterization of SPT order as a form of hidden symmetry breaking to a broad class of models, specifically those for which the SPT order can be related to a \emph{maximally non-commutative} factor system of a finite abelian group that acts on the system through an on-site unitary representation.
Whether any analogous results hold in higher dimensions, or for other kinds of symmetries (e.g.\ time reversal), remains an open question. We point out, however, that in two dimensions and higher there is a different kind of duality that holds for any on-site unitary representation of a finite group $G$: between SPT phases and topological lattice gauge theories with gauge group $G$ \cite{levin_gu,gauge_duality}.

We have also interpreted the action of our generalized KT transformation as a
\emph{topological disentangler}~\cite{topological_disentangler}, removing the
topological component of the entanglement from the ground state.  Transformations that remove entanglement from a quantum many-body system have found use in numerical methods such as the multiscale entanglement renormalisation ansatz (MERA)~\cite{vidal_entanglement_renormalization,vidal_MERA}, and so may the generalized KT transformation presented here.


\begin{acknowledgments}
We acknowledge support from the ARC via the Centre of Excellence in Engineered Quantum Systems (EQuS), project number CE110001013.
\end{acknowledgments}

\appendix

\section{Choice of representative factor system}
\label{choice_of_factor_system}
Here we will discuss the difference between $\mathcal{D}_{\omega}$ and
$\mathcal{D}_{\omega^{\prime}}$, where
\begin{equation}
\omega^{\prime}(g,h) = \beta(g) \beta(h) \beta(gh)^{-1} \omega(g,h) \,,
\end{equation}
[for some set of phase factors $\beta(g)$] is another factor system in the same
cohomology class as $\omega$. The important thing to consider is the transformed
Hamiltonians resulting from the respective transformations.
Thus we will only consider the way
$\mathcal{D}_{\omega}$ acts on symmetry-respecting observables $A$. First of all we
observe that the isomorphism $\varphi_\omega$ only depends on the cohomology
class, so that $\varphi_\omega = \varphi_{\omega^{\prime}}$.
It is then
straightforward to show that
\begin{equation}
\mathcal{D}_{\omega^{\prime}} A \mathcal{D}_{\omega^{\prime}}^{\dagger} =
b_\beta^{\otimes N} \mathcal{D}_\omega A \mathcal{D}^{\dagger}_{\omega}
b_\beta^{\otimes N \dagger},
\end{equation}
where we have defined
\begin{equation}
b_\beta = \sum_{\chi} \beta\bigl(\varphi_\omega^{-1}(\chi)\bigr) \ket{\chi} \bra{\chi}.
\end{equation}
Therefore, the two transformed Hamiltonians differ only be a rephasing of the
basis on each site. Clearly, this does not change the nature of the resulting
transformed phase.

\section{Proof of \eqnref{transformed_string}}
\label{proof_of_transformed_string}
For simplicity of notation, we use the isomorphism $\varphi_\omega$ to 
label our site basis by group elements instead of group characters, i.e.\ we define
\begin{equation}
\ket{g} \equiv \ket{\varphi_\omega(g)}.
\end{equation}
For some choice of character $\chi_{*}$, let $A,B \in \mathcal{A}_{\chi_{*}}$, and define $g_* = \varphi^{-1}_\omega(\chi_{*})$.
This implies that
\begin{align}
  A^\dag & = \sum_{g} \mu_{g} \ket{g} \bra{g g_{*}}
   \\
   B &= \sum_{g} \nu_{g} \ket{g g_{*}} \bra{g}.
\end{align}
for some scalars $\{ \mu_{g} \}$ and $\{ \nu_{g} \}$.
Thus,
\begin{widetext}
\begin{equation}
A^{\dagger} \otimes \mathbb{I}^{\otimes (n-2)} \otimes B 
= \sum_{g_1, \ldots, g_n} \mu_{g_1}
\nu_{g_n}
\ket{g_1, g_2, \ldots, g_n g_*}
\bra{g_1 g_*, \ldots, g_{n-1}, g_n} .
\end{equation}
Therefore 
\begin{equation}
\mathcal{D}_{\omega} (A^{\dagger} \otimes \mathbb{I}^{\otimes (n-2)} \otimes
B) \mathcal{D}_{\omega}^{\dagger} 
= \sum_{g_1, \ldots, g_n} \mu_{g_1}
\nu_{g_n} 
\Gamma_{g_1, \ldots, g_n}
\ket{g_1, g_2, \ldots, g_n g_*}
\bra{g_1 g_*, \ldots, g_{n-1}, g_n} ,
\end{equation}
where $\Gamma_{g_1, \ldots, g_n}$ is the phase factor such
that
\begin{equation}
V_\omega(g_n g_*)
V_\omega(g_{n-1})
\cdots
V_\omega(g_1)= 
\Gamma_{g_1, \ldots, g_n} 
V_\omega(g_n)
\cdots
V_\omega(g_2)
V_\omega(g_1 g_*).
\end{equation}
This gives 
\begin{equation}
\Gamma_{g_1, \ldots, g_n} = 
\omega(g_n, g_{*})^{-1}
\alpha_\omega(g_*, g_{n-1}) \cdots \alpha_\omega(g_*, g_2) \omega(g_*, g_1),
\end{equation}
where $\alpha_\omega(g,h)$ is the phase factor such that
\begin{equation}
V_\omega(g) V_\omega(h) = \alpha_\omega(g,h) V_\omega(h) V_\omega(g).
\end{equation}
However, now comparing with \eqnref{chig}, we find that
\begin{equation}
u(g) \ket{h} = \alpha_\omega(g,h) \ket{h}.
\end{equation}
Therefore, we can conclude that
\begin{equation}
\mathcal{D}_{\omega} (A^\dag \otimes \mathbb{I}^{\otimes (n-2)} \otimes B)
\mathcal{D}_{\omega}^{\dagger} 
= [W_{\omega}(g_*) A^{\dagger}] \otimes [u(g_*)]^{\otimes
(n-2)} \otimes [B W^{\prime}_\omega(g_*)^{\dagger}],
\end{equation}
which leads to \eqnref{transformed_string}, where we have defined
\begin{equation}
  W_\omega(g_*) = \sum_g  \omega(g_*, g) \ket{g} \bra{g} , \qquad
W^{\prime}_\omega(g_*) = \sum_g \omega(g, g_*) \ket{g} \bra{g} .
\end{equation}
\end{widetext}

\section{Identifying phases from patterns of string order}
\label{appendix_pattern_of_string_order}
Symmetry-breaking phases and SPT phases are two different kinds of phases that
can arise in one-dimensional systems invariant under an on-site symmetry.
As was shown in Ref.~\onlinecite{chen_gu_wen_2,schuch}, the most general
kind of phase for such systems combines both aspects. A general symmetric phase for a symmetry group $G$ is characterized by a
subgroup $H$ (corresponding to the symmetries that are unbroken in the bulk) and
a cohomology class $[\omega]$ for $H$, such that each of the degenerate symmetry-breaking
ground states is in the SPT phase $[\omega]$ with respect to the subsymmetry.
In this Appendix, we will show that, in the case of finite
abelian symmetry groups $G$, each distinct phase gives
rise to a distinct pattern of long-range string correlations, as defined through the ``signature''
function $M$ of \eqnref{Mchig}.

Let us first consider the case of pure SPT phases (i.e.\ $H = G$, and the
phases are classified by cohomology classes of $G$). It was argued in
Ref.~\onlinecite{string_order_symmetries} that, if the operators $A$ and $B$ are chosen at
random, then generically one finds that $\langle C_n(A,B;g) \rangle \neq 0$ is
nonzero for any ground state that is invariant under the symmetry in the bulk
(which will be the case for any pure SPT phase as well as the trivial phase).
Suppose, however, that we instead choose $A,B \in
\mathcal{A}_\chi$ for some character $\chi$.  Then, whenever $\chi \neq
\varphi_\omega(g)$, we must be able to find some $g^{\prime} \in G$ such that
$\chi(g^{\prime}) \neq [\varphi_\omega(g)](g^{\prime})$. Recalling the 
  definition of $\varphi_\omega$ [\eqnref{varphi_defn}], and of
$\mathcal{A}_\chi$ [Section \ref{subsec_twoparticle}], this implies that $\alpha_1 \neq
\alpha_2$, where $\alpha_1, \alpha_2$ are the scalars such that
\begin{align}
\label{alpha1}
V_\omega(g) V_\omega(g^{\prime}) &= \alpha_1 V_\omega(g^{\prime}) V_\omega(g),
 \\
A u(g^{\prime}) &= \alpha_2 u(g^{\prime}) A
\label{alpha2}
\end{align}
As shown in Ref.~\onlinecite{detection_string_order}, there is a selection rule that prevents the string correlation $\langle C_n(A,B;g) \rangle$ from being long-ranged when $\alpha_1 \neq \alpha_2$. On the other hand, if $\chi =  \varphi_\omega(g)$ then there is no such selection rule and we expect that $\langle C_n(A,B;g) \rangle$ will generically be long-ranged even with the constraint $A,B \in \mathcal{A}_\chi$. In summary, therefore, the pure SPT phase has the signature $M(\chi,g) = 1 \Leftrightarrow \chi = \varphi_\omega(g)$. For example, 
if the
cohomology class is maximally non-commutative, then $\varphi_\omega$ is invertible and
thus, for each character $\chi$, there is a unique $g$ such that $M(\chi,g) = 1$,
and vice versa. At the other extreme, if the cohomology class is trivial, then
$\varphi_\omega(g) = 1$ for all $g$, and thus $M(\chi,g) = 1 \Leftrightarrow
\chi = 1$.

The above arguments show that the possible signature functions $M$ are in
one-to-one correspondence with the homomorphisms $\varphi_\omega$. In order to
establish that distinct SPT phases correspond to different signature functions,
it only remains to show that if $\omega_1$ and $\omega_2$ have different
cohomology classes, then $\varphi_{\omega_1} \neq \varphi_{\omega_2}$. Since $\varphi_\omega$ is linear in $\omega$, it suffices to prove that if $\varphi_\omega = 1$ (the trivial homomorphism), then $\omega$ has trivial cohomology class. Indeed, $\varphi_\omega = 1$ implies, by definition of $\varphi_\omega$, that $V_\omega(g) V_\omega(g^{\prime}) V_\omega(g)^{\dagger} = V_\omega(g^{\prime})$ for all $g, g^{\prime} \in G$, which is to say all the elements $V_\omega(g)$ commute. If we choose $V_\omega$ to be irreducible, then Schur's Lemma implies that $V_\omega(g) = \beta(g)$ for some scalar phase factors $\beta(g)$. Therefore, the projective representation $V_\omega$ has trivial cohomology class.

Now let us return to the general case, where phases are classified by a subgroup $H \leq
G$ and a cohomology class $[\omega]$ of $H$.
Because all the order parameters we are
considering are expectation values of symmetry-respecting operators,
we just need to determine their value for a single symmetry-breaking ground
state.
Since each of these symmetry-breaking ground states 
lies in an SPT phase with respect to the
subsymmetry $H$, we find that, for $h \in H$, $M(\chi,h) = 1$ if and only if
$\chi_H = \varphi_{\omega}(h)$, where $\chi_H$ is the restriction $\chi$ onto the subgroup $H$.
On the other hand, if $g \notin H$ then $M(\chi,g) = 0$ for any $\chi$
(because the symmetry operation corresponding to $h$ is broken in the bulk \cite{string_order_symmetries}). In summary, the signature of a general
symmetric phase is $M(\chi,g) = 1 \Leftrightarrow [ g \in H \text{ and } \chi_H =
\varphi_\omega(g)].$ Notice that, for $H \neq G$, there will exist non-trivial
characters $\chi$ such that $\chi_H = 1$, which implies that $M(\chi,1) = 1$.
Recall that this corresponds to nonzero values of
\begin{equation}
\lim_{n \to \infty} \langle A^{\dagger} \otimes \mathbb{I}^{\otimes n} \otimes
B
\rangle
\end{equation}
(for $A,B \in \mathcal{A}_{\chi}$),
which is what we expect, since for $H \neq G$ there is partial
symmetry-breaking, and therefore there should also be long-range order.

We will now prove that no two symmetric phases can have the same
signature. Arguing as in the pure SPT case, it is easy to see that for a
\emph{fixed} $H$ all distinct phases have different signatures. To complete the proof, we will now show that the subgroup $H$ can be recovered from the signature, and therefore two phases with different $H$ must have different signatures. To do this we make use of the following result:
\begin{lemma}
\label{lem}
Let $G$ be a finite abelian group, and let $H$ be a subgroup. Then any linear
character acting on $H$ can be extended to a linear character on $G$. That is,
for any $\xi \in H^{*}$, there exists $\chi \in G^{*}$ such that $\chi_H = \xi$.
\begin{proof}
Define the homomorphism $\psi : G^{*} \to H^{*}$, $\chi \mapsto \chi_H$.
Observe that $\ker \psi \cong (G/H)^{*}$, and therefore $|\ker \psi| = |(G/H)^{*}| = |G/H| =
|G|/|H|$. But $\psi(G^{*})
\cong G^{*}/\ker \psi$, so $|\psi(G^{*})| = |G^{*}|/|\ker \psi| = |G|/|\ker
\psi| = |H| = |H^{*}|$. It
follows that $\psi(G^{*}) = H^{*}$, i.e. $\psi$ is surjective.
\end{proof}
\end{lemma}
Lemma \ref{lem} ensures that, for any $h \in H$, we can find a character $\chi
\in G^{*}$ such that $\chi_H = \varphi_{\omega}(h)$, and hence $M(\chi,h)
= 1$. By contrast, if $g \notin H$ then we found above that $M(\chi,g) = 0$ for
all $\chi \in G^{*}$. Therefore, the
subgroup $H$ can be recovered from the signature according to
\begin{equation}
H = \{ h \in G | M(\chi,h) = 1 \text{ for some $\chi \in G^{*}$}\}.
\end{equation}
This completes the proof that distinct phases have distinct signatures.


\begin{thebibliography}{31}%
\makeatletter
\providecommand \@ifxundefined [1]{%
 \@ifx{#1\undefined}
}%
\providecommand \@ifnum [1]{%
 \ifnum #1\expandafter \@firstoftwo
 \else \expandafter \@secondoftwo
 \fi
}%
\providecommand \@ifx [1]{%
 \ifx #1\expandafter \@firstoftwo
 \else \expandafter \@secondoftwo
 \fi
}%
\providecommand \natexlab [1]{#1}%
\providecommand \enquote  [1]{``#1''}%
\providecommand \bibnamefont  [1]{#1}%
\providecommand \bibfnamefont [1]{#1}%
\providecommand \citenamefont [1]{#1}%
\providecommand \href@noop [0]{\@secondoftwo}%
\providecommand \href [0]{\begingroup \@sanitize@url \@href}%
\providecommand \@href[1]{\@@startlink{#1}\@@href}%
\providecommand \@@href[1]{\endgroup#1\@@endlink}%
\providecommand \@sanitize@url [0]{\catcode `\\12\catcode `\$12\catcode
  `\&12\catcode `\#12\catcode `\^12\catcode `\_12\catcode `\%12\relax}%
\providecommand \@@startlink[1]{}%
\providecommand \@@endlink[0]{}%
\providecommand \url  [0]{\begingroup\@sanitize@url \@url }%
\providecommand \@url [1]{\endgroup\@href {#1}{\urlprefix }}%
\providecommand \urlprefix  [0]{URL }%
\providecommand \Eprint [0]{\href }%
\@ifxundefined \urlstyle {%
  \providecommand \doi  [0]{\begingroup \@sanitize@url \@doi}%
  \providecommand \@doi [1]{\endgroup \@@startlink {\doibase
  #1}doi:\discretionary {}{}{}#1\@@endlink }%
}{%
  \providecommand \doi  [0]{doi:\discretionary{}{}{}\begingroup
  \urlstyle{rm}\Url }%
}%
\providecommand \doibase [0]{http://dx.doi.org/}%
\providecommand \Doi [0]{\begingroup \@sanitize@url \@Doi }%
\providecommand \@Doi  [1]{\endgroup\@@startlink{\doibase#1}\@@Doi}%
\providecommand \@@Doi [1]{#1\@@endlink}%
\providecommand \selectlanguage [0]{\@gobble}%
\providecommand \bibinfo  [0]{\@secondoftwo}%
\providecommand \bibfield  [0]{\@secondoftwo}%
\providecommand \translation [1]{[#1]}%
\providecommand \BibitemOpen [0]{}%
\providecommand \bibitemStop [0]{}%
\providecommand \bibitemNoStop [0]{.\EOS\space}%
\providecommand \EOS [0]{\spacefactor3000\relax}%
\providecommand \BibitemShut  [1]{\csname bibitem#1\endcsname}%
\bibitem [{\citenamefont {Chen}\ \emph {et~al.}(2010)\citenamefont {Chen},
  \citenamefont {Gu},\ and\ \citenamefont {Wen}}]{wen_lu}%
  \BibitemOpen
  \bibfield  {author} {\bibinfo {author} {\bibfnamefont {Xie}\ \bibnamefont
  {Chen}}, \bibinfo {author} {\bibfnamefont {Zheng-Cheng}\ \bibnamefont {Gu}},
  \ and\ \bibinfo {author} {\bibfnamefont {Xiao-Gang}\ \bibnamefont {Wen}},\
  }\bibfield  {title} {\enquote {\bibinfo {title} {Local unitary
  transformation, long-range quantum entanglement, wave function
  renormalization, and topological order},}\ }\Doi {10.1103/PhysRevB.82.155138}
  {\bibfield  {journal} {\bibinfo  {journal} {Phys. Rev. B},\ }\textbf
  {\bibinfo {volume} {82}},\ \bibinfo {pages} {155138} (\bibinfo {year}
  {2010})},\ \Eprint {http://arxiv.org/abs/arXiv:1004.3835} {arXiv:1004.3835}
  \BibitemShut {NoStop}%
\bibitem [{\citenamefont {den Nijs}\ and\ \citenamefont
  {Rommelse}(1989)}]{aklt_string_order}%
  \BibitemOpen
  \bibfield  {author} {\bibinfo {author} {\bibfnamefont {Marcel}\ \bibnamefont
  {den Nijs}}\ and\ \bibinfo {author} {\bibfnamefont {Koos}\ \bibnamefont
  {Rommelse}},\ }\bibfield  {title} {\enquote {\bibinfo {title} {Preroughening
  transitions in crystal surfaces and valence-bond phases in quantum spin
  chains},}\ }\Doi {10.1103/PhysRevB.40.4709} {\bibfield  {journal} {\bibinfo
  {journal} {Phys. Rev. B},\ }\textbf {\bibinfo {volume} {40}},\ \bibinfo
  {pages} {4709} (\bibinfo {year} {1989})}\BibitemShut {NoStop}%
\bibitem [{\citenamefont {Polizzi}\ \emph {et~al.}(1998)\citenamefont
  {Polizzi}, \citenamefont {Mila},\ and\ \citenamefont
  {S\o{}rensen}}]{haldane_emergent_edge1}%
  \BibitemOpen
  \bibfield  {author} {\bibinfo {author} {\bibfnamefont {E.}~\bibnamefont
  {Polizzi}}, \bibinfo {author} {\bibfnamefont {F.}~\bibnamefont {Mila}}, \
  and\ \bibinfo {author} {\bibfnamefont {E.~S.}\ \bibnamefont {S\o{}rensen}},\
  }\bibfield  {title} {\enquote {\bibinfo {title} {{$S=1/2$} chain-boundary
  excitations in the {Haldane} phase of one-dimensional {$S=1$} systems},}\
  }\Doi {10.1103/PhysRevB.58.2407} {\bibfield  {journal} {\bibinfo  {journal}
  {Phys. Rev. B},\ }\textbf {\bibinfo {volume} {58}},\ \bibinfo {pages} {2407}
  (\bibinfo {year} {1998})}\BibitemShut {NoStop}%
\bibitem [{\citenamefont {Pollmann}\ \emph {et~al.}(2010)\citenamefont
  {Pollmann}, \citenamefont {Turner}, \citenamefont {Berg},\ and\ \citenamefont
  {Oshikawa}}]{pollmann-prb-2010}%
  \BibitemOpen
  \bibfield  {author} {\bibinfo {author} {\bibfnamefont {Frank}\ \bibnamefont
  {Pollmann}}, \bibinfo {author} {\bibfnamefont {Ari~M.}\ \bibnamefont
  {Turner}}, \bibinfo {author} {\bibfnamefont {Erez}\ \bibnamefont {Berg}}, \
  and\ \bibinfo {author} {\bibfnamefont {Masaki}\ \bibnamefont {Oshikawa}},\
  }\bibfield  {title} {\enquote {\bibinfo {title} {Entanglement spectrum of a
  topological phase in one dimension},}\ }\Doi {10.1103/PhysRevB.81.064439}
  {\bibfield  {journal} {\bibinfo  {journal} {Phys. Rev. B},\ }\textbf
  {\bibinfo {volume} {81}},\ \bibinfo {pages} {064439} (\bibinfo {year}
  {2010})},\ \Eprint {http://arxiv.org/abs/arXiv:0910.1811} {arXiv:0910.1811}
  \BibitemShut {NoStop}%
\bibitem [{\citenamefont {Verstraete}\ \emph {et~al.}(2004)\citenamefont
  {Verstraete}, \citenamefont {Mart\'in-Delgado},\ and\ \citenamefont
  {Cirac}}]{diverging_prl}%
  \BibitemOpen
  \bibfield  {author} {\bibinfo {author} {\bibfnamefont {F.}~\bibnamefont
  {Verstraete}}, \bibinfo {author} {\bibfnamefont {M.~A.}\ \bibnamefont
  {Mart\'in-Delgado}}, \ and\ \bibinfo {author} {\bibfnamefont {J.~I.}\
  \bibnamefont {Cirac}},\ }\bibfield  {title} {\enquote {\bibinfo {title}
  {Diverging entanglement length in gapped quantum spin systems},}\ }\Doi
  {10.1103/PhysRevLett.92.087201} {\bibfield  {journal} {\bibinfo  {journal}
  {Phys. Rev. Lett.},\ }\textbf {\bibinfo {volume} {92}},\ \bibinfo {pages}
  {087201} (\bibinfo {year} {2004})},\ \Eprint {http://arxiv.org/abs/0311087}
  {arXiv:0311087} \BibitemShut {NoStop}%
\bibitem [{\citenamefont {Venuti}\ and\ \citenamefont
  {Roncaglia}(2005)}]{camposvenuti}%
  \BibitemOpen
  \bibfield  {author} {\bibinfo {author} {\bibfnamefont {L.~Campos}\
  \bibnamefont {Venuti}}\ and\ \bibinfo {author} {\bibfnamefont
  {M.}~\bibnamefont {Roncaglia}},\ }\bibfield  {title} {\enquote {\bibinfo
  {title} {Analytic relations between localizable entanglement and string
  correlations in spin systems},}\ }\Doi {10.1103/PhysRevLett.94.207207}
  {\bibfield  {journal} {\bibinfo  {journal} {Phys. Rev. Lett.},\ }\textbf
  {\bibinfo {volume} {94}},\ \bibinfo {pages} {207207} (\bibinfo {year}
  {2005})},\ \Eprint {http://arxiv.org/abs/cond-mat/0503021}
  {arXiv:cond-mat/0503021} \BibitemShut {NoStop}%
\bibitem [{\citenamefont {Chen}\ \emph
  {et~al.}(2011){\natexlab{a}}\citenamefont {Chen}, \citenamefont {Gu},\ and\
  \citenamefont {Wen}}]{chen_gu_wen}%
  \BibitemOpen
  \bibfield  {author} {\bibinfo {author} {\bibfnamefont {Xie}\ \bibnamefont
  {Chen}}, \bibinfo {author} {\bibfnamefont {Zheng-Cheng}\ \bibnamefont {Gu}},
  \ and\ \bibinfo {author} {\bibfnamefont {Xiao-Gang}\ \bibnamefont {Wen}},\
  }\bibfield  {title} {\enquote {\bibinfo {title} {Classification of gapped
  symmetric phases in one-dimensional spin systems},}\ }\Doi
  {10.1103/PhysRevB.83.035107} {\bibfield  {journal} {\bibinfo  {journal}
  {Phys. Rev. B},\ }\textbf {\bibinfo {volume} {83}},\ \bibinfo {pages}
  {035107} (\bibinfo {year} {2011}{\natexlab{a}})},\ \Eprint
  {http://arxiv.org/abs/arXiv:1008.3745} {arXiv:1008.3745} \BibitemShut
  {NoStop}%
\bibitem [{\citenamefont {Schuch}\ \emph {et~al.}(2011)\citenamefont {Schuch},
  \citenamefont {P{\'e}rez-Garc{\'i}a},\ and\ \citenamefont {Cirac}}]{schuch}%
  \BibitemOpen
  \bibfield  {author} {\bibinfo {author} {\bibfnamefont {Norbert}\ \bibnamefont
  {Schuch}}, \bibinfo {author} {\bibfnamefont {David}\ \bibnamefont
  {P{\'e}rez-Garc{\'i}a}}, \ and\ \bibinfo {author} {\bibfnamefont {Ignacio}\
  \bibnamefont {Cirac}},\ }\bibfield  {title} {\enquote {\bibinfo {title}
  {Classifying quantum phases using matrix product states and projected
  entangled pair states},}\ }\Doi {10.1103/PhysRevB.84.165139} {\bibfield
  {journal} {\bibinfo  {journal} {Phys. Rev. B},\ }\textbf {\bibinfo {volume}
  {84}},\ \bibinfo {pages} {165139} (\bibinfo {year} {2011})},\ \Eprint
  {http://arxiv.org/abs/1010.3732} {arXiv:1010.3732} \BibitemShut {NoStop}%
\bibitem [{\citenamefont {Kennedy}\ and\ \citenamefont
  {Tasaki}(1992){\natexlab{a}}}]{kt}%
  \BibitemOpen
  \bibfield  {author} {\bibinfo {author} {\bibfnamefont {Tom}\ \bibnamefont
  {Kennedy}}\ and\ \bibinfo {author} {\bibfnamefont {Hal}\ \bibnamefont
  {Tasaki}},\ }\bibfield  {title} {\enquote {\bibinfo {title} {Hidden {$Z_2
  \times Z_2$} symmetry breaking in {Haldane}-gap antiferromagnets},}\ }\Doi
  {10.1103/PhysRevB.45.304} {\bibfield  {journal} {\bibinfo  {journal} {Phys.
  Rev. B},\ }\textbf {\bibinfo {volume} {45}},\ \bibinfo {pages} {304--307}
  (\bibinfo {year} {1992}{\natexlab{a}})}\BibitemShut {NoStop}%
\bibitem [{\citenamefont {Kennedy}\ and\ \citenamefont
  {Tasaki}(1992){\natexlab{b}}}]{kt2}%
  \BibitemOpen
  \bibfield  {author} {\bibinfo {author} {\bibfnamefont {Tom}\ \bibnamefont
  {Kennedy}}\ and\ \bibinfo {author} {\bibfnamefont {Hal}\ \bibnamefont
  {Tasaki}},\ }\bibfield  {title} {\enquote {\bibinfo {title} {Hidden symmetry
  breaking and the {Haldane} phase in {$S = 1$} quantum spin chains},}\ }\Doi
  {10.1007/BF02097239} {\bibfield  {journal} {\bibinfo  {journal} {Comm. Math.
  Phys.},\ }\textbf {\bibinfo {volume} {147}},\ \bibinfo {pages} {431}
  (\bibinfo {year} {1992}{\natexlab{b}})}\BibitemShut {NoStop}%
\bibitem [{\citenamefont {Pollmann}\ \emph {et~al.}(2012)\citenamefont
  {Pollmann}, \citenamefont {Berg}, \citenamefont {Turner},\ and\ \citenamefont
  {Oshikawa}}]{pollmann-arxiv-2009}%
  \BibitemOpen
  \bibfield  {author} {\bibinfo {author} {\bibfnamefont {Frank}\ \bibnamefont
  {Pollmann}}, \bibinfo {author} {\bibfnamefont {Erez}\ \bibnamefont {Berg}},
  \bibinfo {author} {\bibfnamefont {Ari~M.}\ \bibnamefont {Turner}}, \ and\
  \bibinfo {author} {\bibfnamefont {Masaki}\ \bibnamefont {Oshikawa}},\
  }\bibfield  {title} {\enquote {\bibinfo {title} {Symmetry protection of
  topological phases in one-dimensional quantum spin systems},}\ }\Doi
  {10.1103/PhysRevB.85.075125} {\bibfield  {journal} {\bibinfo  {journal}
  {Phys. Rev. B},\ }\textbf {\bibinfo {volume} {85}},\ \bibinfo {pages}
  {075125} (\bibinfo {year} {2012})},\ \Eprint {http://arxiv.org/abs/0909.4059}
  {arXiv:0909.4059} \BibitemShut {NoStop}%
\bibitem [{\citenamefont {Tu}\ \emph {et~al.}(2008)\citenamefont {Tu},
  \citenamefont {Zhang},\ and\ \citenamefont {Xiang}}]{SO_string_order}%
  \BibitemOpen
  \bibfield  {author} {\bibinfo {author} {\bibfnamefont {Hong-Hao}\
  \bibnamefont {Tu}}, \bibinfo {author} {\bibfnamefont {Guang-Ming}\
  \bibnamefont {Zhang}}, \ and\ \bibinfo {author} {\bibfnamefont {Tao}\
  \bibnamefont {Xiang}},\ }\bibfield  {title} {\enquote {\bibinfo {title}
  {String order and hidden topological symmetry in the {$\mathrm{SO}(2n + 1)$}
  symmetric matrix product states},}\ }\Doi {10.1088/1751-8113/41/41/415201}
  {\bibfield  {journal} {\bibinfo  {journal} {J. Phys. A: Math. Theor.},\
  }\textbf {\bibinfo {volume} {41}},\ \bibinfo {pages} {415201} (\bibinfo
  {year} {2008})},\ \Eprint {http://arxiv.org/abs/0804.1685} {arXiv:0804.1685}
  \BibitemShut {NoStop}%
\bibitem [{\citenamefont {Else}\ \emph
  {et~al.}(2012){\natexlab{a}}\citenamefont {Else}, \citenamefont {Schwarz},
  \citenamefont {Bartlett},\ and\ \citenamefont
  {Doherty}}]{else_schwarz_bartlett_doherty_symmetry}%
  \BibitemOpen
  \bibfield  {author} {\bibinfo {author} {\bibfnamefont {Dominic~V.}\
  \bibnamefont {Else}}, \bibinfo {author} {\bibfnamefont {Ilai}\ \bibnamefont
  {Schwarz}}, \bibinfo {author} {\bibfnamefont {Stephen~D.}\ \bibnamefont
  {Bartlett}}, \ and\ \bibinfo {author} {\bibfnamefont {Andrew~C.}\
  \bibnamefont {Doherty}},\ }\bibfield  {title} {\enquote {\bibinfo {title}
  {Symmetry-protected phases for measurement-based quantum computation},}\
  }\Doi {10.1103/PhysRevLett.108.240505} {\bibfield  {journal} {\bibinfo
  {journal} {Phys. Rev. Lett.},\ }\textbf {\bibinfo {volume} {108}},\ \bibinfo
  {pages} {240505} (\bibinfo {year} {2012}{\natexlab{a}})},\ \Eprint
  {http://arxiv.org/abs/1201.4877} {arXiv:1201.4877} \BibitemShut {NoStop}%
\bibitem [{\citenamefont {Else}\ \emph
  {et~al.}(2012){\natexlab{b}}\citenamefont {Else}, \citenamefont {Bartlett},\
  and\ \citenamefont {Doherty}}]{big_paper}%
  \BibitemOpen
  \bibfield  {author} {\bibinfo {author} {\bibfnamefont {Dominic~V.}\
  \bibnamefont {Else}}, \bibinfo {author} {\bibfnamefont {Stephen~D.}\
  \bibnamefont {Bartlett}}, \ and\ \bibinfo {author} {\bibfnamefont
  {Andrew~C.}\ \bibnamefont {Doherty}},\ }\bibfield  {title} {\enquote
  {\bibinfo {title} {Symmetry protection of measurement-based quantum
  computation in ground states},}\ }\href@noop {} {\bibfield  {journal}
  {\bibinfo  {journal} {New J. Phys.},\ }\textbf {\bibinfo {volume} {14}},\
  \bibinfo {pages} {113016} (\bibinfo {year} {2012}{\natexlab{b}})},\ \Eprint
  {http://arxiv.org/abs/1207.4805} {arXiv:1207.4805} \BibitemShut {NoStop}%
\bibitem [{\citenamefont {Pollmann}\ and\ \citenamefont
  {Turner}(2012)}]{detection_string_order}%
  \BibitemOpen
  \bibfield  {author} {\bibinfo {author} {\bibfnamefont {Frank}\ \bibnamefont
  {Pollmann}}\ and\ \bibinfo {author} {\bibfnamefont {Ari~M.}\ \bibnamefont
  {Turner}},\ }\bibfield  {title} {\enquote {\bibinfo {title} {Detection of
  symmetry-protected topological phases in one dimension},}\ }\Doi
  {10.1103/PhysRevB.86.125441} {\bibfield  {journal} {\bibinfo  {journal}
  {Phys. Rev. B},\ }\textbf {\bibinfo {volume} {86}},\ \bibinfo {pages}
  {125441} (\bibinfo {year} {2012})},\ \Eprint {http://arxiv.org/abs/1204.0704}
  {arXiv:1204.0704} \BibitemShut {NoStop}%
\bibitem [{\citenamefont {Duivenvoorden}\ and\ \citenamefont
  {Quella}(2013)}]{duivenvoorden_quella}%
  \BibitemOpen
  \bibfield  {author} {\bibinfo {author} {\bibfnamefont {Kasper}\ \bibnamefont
  {Duivenvoorden}}\ and\ \bibinfo {author} {\bibfnamefont {Thomas}\
  \bibnamefont {Quella}},\ }\href@noop {} {\enquote {\bibinfo {title} {From
  symmetry-protected topological order to landau order},}\ } (\bibinfo {year}
  {2013}),\ \Eprint {http://arxiv.org/abs/1304.7234} {arXiv:1304.7234}
  \BibitemShut {NoStop}%
\bibitem [{\citenamefont {Oshikawa}(1992)}]{kt_arbitrary_spin}%
  \BibitemOpen
  \bibfield  {author} {\bibinfo {author} {\bibfnamefont {M}~\bibnamefont
  {Oshikawa}},\ }\bibfield  {title} {\enquote {\bibinfo {title} {Hidden {$Z_2
  \times Z_2$} symmetry in quantum spin chains with arbitrary integer spin},}\
  }\Doi {10.1088/0953-8984/4/36/019} {\bibfield  {journal} {\bibinfo  {journal}
  {J. Phys.: Condens. Matter},\ }\textbf {\bibinfo {volume} {4}},\ \bibinfo
  {pages} {7469} (\bibinfo {year} {1992})}\BibitemShut {NoStop}%
\bibitem [{\citenamefont {Berkovich}\ and\ \citenamefont
  {Zhmud${}^\prime$}(1998)}]{berkovich_characters}%
  \BibitemOpen
  \bibfield  {author} {\bibinfo {author} {\bibfnamefont {Ya.~G.}\ \bibnamefont
  {Berkovich}}\ and\ \bibinfo {author} {\bibfnamefont {E.~M.}\ \bibnamefont
  {Zhmud${}^\prime$}},\ }\href@noop {} {\emph {\bibinfo {title} {Characters of
  Finite Groups}}},\ Vol.~\bibinfo {volume} {1}\ (\bibinfo  {publisher}
  {American Mathematical Society},\ \bibinfo {address} {Providence, Rhode
  Island},\ \bibinfo {year} {1998})\BibitemShut {NoStop}%
\bibitem [{\citenamefont {Frucht}(1932)}]{frucht1932}%
  \BibitemOpen
  \bibfield  {author} {\bibinfo {author} {\bibfnamefont {R.}~\bibnamefont
  {Frucht}},\ }\bibfield  {title} {\enquote {\bibinfo {title} {{\"U}ber die
  darstellung endlicher abelscher gruppen durch kollineationen},}\ }\Doi
  {10.1515/crll.1932.166.16} {\bibfield  {journal} {\bibinfo  {journal}
  {J. Reine Angew. Math.},\ }\textbf {\bibinfo
  {volume} {1932}},\ \bibinfo {pages} {16} (\bibinfo {year}
  {1932})}\BibitemShut {NoStop}%
\bibitem [{\citenamefont {Affleck}\ \emph {et~al.}(1988)\citenamefont
  {Affleck}, \citenamefont {Kennedy}, \citenamefont {Lieb},\ and\ \citenamefont
  {Tasaki}}]{aklt1}%
  \BibitemOpen
  \bibfield  {author} {\bibinfo {author} {\bibfnamefont {I.}~\bibnamefont
  {Affleck}}, \bibinfo {author} {\bibfnamefont {T.}~\bibnamefont {Kennedy}},
  \bibinfo {author} {\bibfnamefont {E.H.}\ \bibnamefont {Lieb}}, \ and\
  \bibinfo {author} {\bibfnamefont {H.}~\bibnamefont {Tasaki}},\ }\bibfield
  {title} {\enquote {\bibinfo {title} {Valence bond ground states in isotropic
  quantum antiferromagnets},}\ }\Doi {10.1007/BF01218021} {\bibfield  {journal}
  {\bibinfo  {journal} {Comm. Math. Phys.},\ }\textbf {\bibinfo {volume}
  {115}},\ \bibinfo {pages} {477} (\bibinfo {year} {1988})}\BibitemShut
  {NoStop}%
\bibitem [{\citenamefont {Affleck}\ \emph {et~al.}(1987)\citenamefont
  {Affleck}, \citenamefont {Kennedy}, \citenamefont {Lieb},\ and\ \citenamefont
  {Tasaki}}]{aklt2}%
  \BibitemOpen
  \bibfield  {author} {\bibinfo {author} {\bibfnamefont {Ian}\ \bibnamefont
  {Affleck}}, \bibinfo {author} {\bibfnamefont {Tom}\ \bibnamefont {Kennedy}},
  \bibinfo {author} {\bibfnamefont {Elliott~H.}\ \bibnamefont {Lieb}}, \ and\
  \bibinfo {author} {\bibfnamefont {Hal}\ \bibnamefont {Tasaki}},\ }\bibfield
  {title} {\enquote {\bibinfo {title} {Rigorous results on valence-bond ground
  states in antiferromagnets},}\ }\Doi {10.1103/PhysRevLett.59.799} {\bibfield
  {journal} {\bibinfo  {journal} {Phys. Rev. Lett.},\ }\textbf {\bibinfo
  {volume} {59}},\ \bibinfo {pages} {799} (\bibinfo {year} {1987})}\BibitemShut
  {NoStop}%
\bibitem [{\citenamefont {Perez-Garcia}\ \emph {et~al.}(2007)\citenamefont
  {Perez-Garcia}, \citenamefont {Verstraete}, \citenamefont {Wolf},\ and\
  \citenamefont {Cirac}}]{perezgarcia_mps}%
  \BibitemOpen
  \bibfield  {author} {\bibinfo {author} {\bibfnamefont {D.}~\bibnamefont
  {Perez-Garcia}}, \bibinfo {author} {\bibfnamefont {F.}~\bibnamefont
  {Verstraete}}, \bibinfo {author} {\bibfnamefont {M.~M.}\ \bibnamefont
  {Wolf}}, \ and\ \bibinfo {author} {\bibfnamefont {J.~I.}\ \bibnamefont
  {Cirac}},\ }\bibfield  {title} {\enquote {\bibinfo {title} {Matrix product
  state representations},}\ }\href@noop {} {\bibfield  {journal} {\bibinfo
  {journal} {Quant.\ Inf.\ Comput.},\ }\textbf {\bibinfo {volume} {7}},\
  \bibinfo {pages} {401} (\bibinfo {year} {2007})},\ \Eprint
  {http://arxiv.org/abs/arXiv:quant-ph/0608197} {arXiv:quant-ph/0608197}
  \BibitemShut {NoStop}%
\bibitem [{\citenamefont {P\'erez-Garc\'\i{}a}\ \emph
  {et~al.}(2008)\citenamefont {P\'erez-Garc\'\i{}a}, \citenamefont {Wolf},
  \citenamefont {Sanz}, \citenamefont {Verstraete},\ and\ \citenamefont
  {Cirac}}]{string_order_symmetries}%
  \BibitemOpen
  \bibfield  {author} {\bibinfo {author} {\bibfnamefont {D.}~\bibnamefont
  {P\'erez-Garc\'\i{}a}}, \bibinfo {author} {\bibfnamefont {M.~M.}\
  \bibnamefont {Wolf}}, \bibinfo {author} {\bibfnamefont {M.}~\bibnamefont
  {Sanz}}, \bibinfo {author} {\bibfnamefont {F.}~\bibnamefont {Verstraete}}, \
  and\ \bibinfo {author} {\bibfnamefont {J.~I.}\ \bibnamefont {Cirac}},\
  }\bibfield  {title} {\enquote {\bibinfo {title} {String order and symmetries
  in quantum spin lattices},}\ }\Doi {10.1103/PhysRevLett.100.167202}
  {\bibfield  {journal} {\bibinfo  {journal} {Phys. Rev. Lett.},\ }\textbf
  {\bibinfo {volume} {100}},\ \bibinfo {pages} {167202} (\bibinfo {year}
  {2008})},\ \Eprint {http://arxiv.org/abs/arXiv:0802.0447} {arXiv:0802.0447}
  \BibitemShut {NoStop}%
\bibitem [{\citenamefont {Chen}\ \emph
  {et~al.}(2011){\natexlab{b}}\citenamefont {Chen}, \citenamefont {Gu},\ and\
  \citenamefont {Wen}}]{chen_gu_wen_2}%
  \BibitemOpen
  \bibfield  {author} {\bibinfo {author} {\bibfnamefont {Xie}\ \bibnamefont
  {Chen}}, \bibinfo {author} {\bibfnamefont {Zheng-Cheng}\ \bibnamefont {Gu}},
  \ and\ \bibinfo {author} {\bibfnamefont {Xiao-Gang}\ \bibnamefont {Wen}},\
  }\bibfield  {title} {\enquote {\bibinfo {title} {Complete classification of
  one-dimensional gapped quantum phases in interacting spin systems},}\ }\Doi
  {10.1103/PhysRevB.84.235128} {\bibfield  {journal} {\bibinfo  {journal}
  {Phys. Rev. B},\ }\textbf {\bibinfo {volume} {84}},\ \bibinfo {pages}
  {235128} (\bibinfo {year} {2011}{\natexlab{b}})},\ \Eprint
  {http://arxiv.org/abs/1103.3323} {arXiv:1103.3323} \BibitemShut {NoStop}%
\bibitem [{\citenamefont {Duivenvoorden}\ and\ \citenamefont
  {Quella}(2012)}]{suN_phases}%
  \BibitemOpen
  \bibfield  {author} {\bibinfo {author} {\bibfnamefont {Kasper}\ \bibnamefont
  {Duivenvoorden}}\ and\ \bibinfo {author} {\bibfnamefont {Thomas}\
  \bibnamefont {Quella}},\ }
  \bibfield  {title} {\enquote {\bibinfo {title} 
  {On topological phases of spin chains},}\ }\Doi
  {10.1103/PhysRevB.87.125145} {\bibfield  {journal} {\bibinfo  {journal}
  {Phys. Rev. B},\ }\textbf {\bibinfo {volume} {87}},\ \bibinfo {pages}
  {125145} (\bibinfo {year} {2013})},\ \Eprint
  {http://arxiv.org/abs/1206.2462} {arXiv:1206.2462} \BibitemShut {NoStop}%
\bibitem [{\citenamefont {Verstraete}\ \emph {et~al.}(2005)\citenamefont
  {Verstraete}, \citenamefont {Cirac}, \citenamefont {Latorre}, \citenamefont
  {Rico},\ and\ \citenamefont {Wolf}}]{verstraete}%
  \BibitemOpen
  \bibfield  {author} {\bibinfo {author} {\bibfnamefont {F.}~\bibnamefont
  {Verstraete}}, \bibinfo {author} {\bibfnamefont {J.~I.}\ \bibnamefont
  {Cirac}}, \bibinfo {author} {\bibfnamefont {J.~I.}\ \bibnamefont {Latorre}},
  \bibinfo {author} {\bibfnamefont {E.}~\bibnamefont {Rico}}, \ and\ \bibinfo
  {author} {\bibfnamefont {M.~M.}\ \bibnamefont {Wolf}},\ }\bibfield  {title}
  {\enquote {\bibinfo {title} {Renormalization-group transformations on quantum
  states},}\ }\Doi {10.1103/PhysRevLett.94.140601} {\bibfield  {journal}
  {\bibinfo  {journal} {Phys. Rev. Lett.},\ }\textbf {\bibinfo {volume} {94}},\
  \bibinfo {pages} {140601} (\bibinfo {year} {2005})},\ \Eprint
  {http://arxiv.org/abs/arXiv:quant-ph/0410227} {arXiv:quant-ph/0410227}
  \BibitemShut {NoStop}%
\bibitem [{\citenamefont {Okunishi}(2011)}]{topological_disentangler}%
  \BibitemOpen
  \bibfield  {author} {\bibinfo {author} {\bibfnamefont {Kouichi}\ \bibnamefont
  {Okunishi}},\ }\bibfield  {title} {\enquote {\bibinfo {title} {Topological
  disentangler for the valence-bond-solid chain},}\ }\Doi
  {10.1103/PhysRevB.83.104411} {\bibfield  {journal} {\bibinfo  {journal}
  {Phys. Rev. B},\ }\textbf {\bibinfo {volume} {83}},\ \bibinfo {pages}
  {104411} (\bibinfo {year} {2011})},\ \Eprint {http://arxiv.org/abs/1011.3277}
  {arXiv:1011.3277} \BibitemShut {NoStop}%
\bibitem [{\citenamefont {Levin}\ and\ \citenamefont {Gu}(2012)}]{levin_gu}%
  \BibitemOpen
  \bibfield  {author} {\bibinfo {author} {\bibfnamefont {Michael}\ \bibnamefont
  {Levin}}\ and\ \bibinfo {author} {\bibfnamefont {Zheng-Cheng}\ \bibnamefont
  {Gu}},\ }\bibfield  {title} {\enquote {\bibinfo {title} {Braiding statistics
  approach to symmetry-protected topological phases},}\ }\Doi
  {10.1103/PhysRevB.86.115109} {\bibfield  {journal} {\bibinfo  {journal}
  {Phys. Rev. B},\ }\textbf {\bibinfo {volume} {86}},\ \bibinfo {pages}
  {115109} (\bibinfo {year} {2012})},\ \Eprint {http://arxiv.org/abs/1202.3120}
  {arXiv:1202.3120} \BibitemShut {NoStop}%
\bibitem [{\citenamefont {Hung}\ and\ \citenamefont
  {Wen}(2012)}]{gauge_duality}%
  \BibitemOpen
  \bibfield  {author} {\bibinfo {author} {\bibfnamefont {Ling-Yang}\
  \bibnamefont {Hung}}\ and\ \bibinfo {author} {\bibfnamefont {Xiao-Gang}\
  \bibnamefont {Wen}},\ }\href@noop {} {\enquote {\bibinfo {title} {Quantized
  topological terms in weakly coupled gauge theories and their connection to
  symmetry protected topological phases},}\ } (\bibinfo {year} {2012}),\
  \Eprint {http://arxiv.org/abs/1211.2767} {arXiv:1211.2767} \BibitemShut
  {NoStop}%
\bibitem [{\citenamefont {Vidal}(2007)}]{vidal_entanglement_renormalization}%
  \BibitemOpen
  \bibfield  {author} {\bibinfo {author} {\bibfnamefont {G.}~\bibnamefont
  {Vidal}},\ }\bibfield  {title} {\enquote {\bibinfo {title} {Entanglement
  renormalization},}\ }\Doi {10.1103/PhysRevLett.99.220405} {\bibfield
  {journal} {\bibinfo  {journal} {Phys. Rev. Lett.},\ }\textbf {\bibinfo
  {volume} {99}},\ \bibinfo {pages} {220405} (\bibinfo {year} {2007})},\
  \Eprint {http://arxiv.org/abs/arXiv:cond-mat/0512165}
  {arXiv:cond-mat/0512165} \BibitemShut {NoStop}%
\bibitem [{\citenamefont {Vidal}(2008)}]{vidal_MERA}%
  \BibitemOpen
  \bibfield  {author} {\bibinfo {author} {\bibfnamefont {G.}~\bibnamefont
  {Vidal}},\ }\bibfield  {title} {\enquote {\bibinfo {title} {Class of quantum
  many-body states that can be efficiently simulated},}\ }\Doi
  {10.1103/PhysRevLett.101.110501} {\bibfield  {journal} {\bibinfo  {journal}
  {Phys. Rev. Lett.},\ }\textbf {\bibinfo {volume} {101}},\ \bibinfo {pages}
  {110501} (\bibinfo {year} {2008})},\ \Eprint
  {http://arxiv.org/abs/arXiv:quant-ph/0610099} {arXiv:quant-ph/0610099}
  \BibitemShut {NoStop}%
\end{thebibliography}
\end{document}